\documentclass[twocolumn]{aastex63}
\pdfoutput=1

\usepackage{amsmath}
\usepackage{hyperref}

\newcommand{\rev}[1]{\textcolor{black}{ #1}}
\newcommand{\nrev}[1]{\textcolor{black}{ #1}}

\definecolor{light-gray}{gray}{0.8}
\newcommand{\V}[1]{\mathbf{#1}}

\revised{...}
\accepted{...}
\submitjournal{ApJL}

\shorttitle{Relativistic Turbulence}
\shortauthors{Chernoglazov, A. et al.}

\graphicspath{{./}{figures/}}

\begin{document}

\title{Dynamic alignment and plasmoid formation in relativistic magnetohydrodynamic turbulence}

\correspondingauthor{Alexander Chernoglazov}
\email{alexander.chernoglazov@gmail.com}

\author{Alexander Chernoglazov}
\affiliation{Department of Physics, University of New Hampshire, 9 Library Way, Durham NH 03824, USA}
\affiliation{Center for Computational Astrophysics, Flatiron Institute, 162 Fifth Avenue, New York, NY 10010, USA}

\author[0000-0002-7301-3908]{Bart Ripperda}
\altaffiliation{Joint Princeton/Flatiron Postdoctoral Fellow}\affiliation{Center for Computational Astrophysics, Flatiron Institute, 162 Fifth Avenue, New York, NY 10010, USA}\affiliation{Department of Astrophysical Sciences, Peyton Hall, Princeton University, Princeton, NJ 08544, USA}

\author[0000-0001-7801-0362]{Alexander Philippov}
\affiliation{Center for Computational Astrophysics, Flatiron Institute, 162 Fifth Avenue, New York, NY 10010, USA}

\begin{abstract}

We present high resolution 2D and 3D simulations of magnetized decaying turbulence in relativistic resistive magneto-hydrodynamics. The simulations show dynamic formation of large scale intermittent long-lived current sheets being disrupted by the tearing instability into plasmoid chains. These current sheets are locations of enhanced magnetic field dissipation and heating of the plasma. \nrev{We find magnetic energy spectra $\propto k^{-3/2}$, together with strongly pronounced dynamic alignment of Elsasser fields and of velocity and magnetic fields, for strong guide-field turbulence, whereas we retrieve spectra $\propto k^{-5/3}$ for the case of a weak guide-field.}
%We determine the \nrev{formations of the magnetic energy spectrum $k^{-3/2}$ for the most of simulations instead of $k^{-5/3}$ and dynamic alignment of the velocity and magnetic fields and Elsasser fields in both 2D and 3D relativistic turbulence. The review of different prediction of the dynamic alignment formations is presented.} \crossout{of demonstrate full agreement with the theoretical prediction of $r^{1/4}$} in the inertial range. 
\end{abstract}


\section{Introduction} \label{sec:intro}

Turbulence provides a route for the energy cascade and dissipation in a wide range of astrophysical plasmas. This is relevant for astrophysical systems like black hole accretion disk-jet systems \citep[e.g.,][]{Ripperda_2020,RipperdaFlares, Mahlmann_2020}, magnetar magnetospheres \citep{beloborodov2020emission} and pulsar wind nebulae \citep[e.g.,][]{Lyubarsky1992,Begelman1998}. 
\rev{These astrophysical systems are typically relativistic, meaning that the magnetization \mbox{$\sigma = B^2 / (4\pi \omega \rho c^2) \geq 1$}, where $B$ is the magnetic field strength, $\rho$ is the plasma density, and $\omega$ is the relativistic enthalpy density, indicating that the magnetic energy density is larger than the plasma energy density. This results in an Alfv\'{e}n speed $v_{\rm A} = \sqrt{\sigma / (\sigma+1)}c$ that is close to the speed of light $c$.}
%\rev{All of these systems share some common features: plasmas there are in relativistic state, meaning Alfv\'{e}n speed $v_a$ is close to the speed of light $c$; they have large scale strong magnetic field ${\bf B}_0$; perturbations of this field $\delta {\bf B}$ can reach comparable amplitude for the most energetic phenomena.}

Most turbulence studies have been in the realm of non-relativistic magnetohydrodynamics (MHD) when the Alfv\'{e}n speed, $v_A$, is much lower than the speed of light, $c$. \cite{Iroshnikov1963,Kraichnan1965} showed that the energy cascade from large to small scales is caused by the mutual shear of counter-propagating Alfv\'{e}n waves. Thirty years later \cite{GS95,GS_revisited} suggested that  turbulent systems are in the critical balance regime meaning that an eddy \rev{is} significantly deform\rev{ed} during one Alfv\'{e}n-crossing time. This also means that the turbulent eddies are elongated along the background magnetic field. The first steps towards a theory of relativistic turbulence were taken recently by \cite{Chandran_2018}, and they demonstrated that the relativistic picture is very similar to the Newtonian limit (\rev{more details are presented in Section \ref{sec:mhdcomparison}).} \cite{Boldyrev_2005,Boldyrev_2006} suggested that turbulent eddies are anisotropic in all three directions: they are elongated along the guide magnetic field and have two different sizes in the guide field-perpendicular plane. The ratio of these two sizes is called dynamic alignment \rev{angle}. These eddies are progressively more elongated at smaller scales. Recent theories \citep[e.g.,][]{Boldyrev_Loureiro,Mallet_2017} proposed that the elongated eddies at small enough scale become unstable to the tearing instability, causing a steepening of the turbulence spectrum. 

In their recent paper, \citealt{Dong2018} demonstrated the formation of reconnecting current sheets in two-dimensional (2D) decaying non-relativistic turbulence. They also demonstrated the formation of a turbulence spectrum and dynamic alignment in agreement with Boldyrev's theory. It is however as of yet unclear whether these findings persist in the case of realistic three-dimensional (3D) turbulence. In 3D MHD, despite prominent current sheet formation \citep{Zhdankin}, it remains unclear whether reconnection can occur in the fast regime, when the dissipation efficiency is independent of resistivity.  This regime is associated with the formation of plasmoid chains, \rev{resulting in a universal reconnection rate of order $0.01$ (\citealt{bhattacharjee2009}; \citealt{uzdensky2010}).}
%\rev{ and maintains reconnnection rate $v_{\rm in}/v_{\rm A}$ of order $10^{-2}$ (\citealt{bhattacharjee2009}; \citealt{uzdensky2010}).}\crossout{\citep{Loureiro2007}} . 

\rev{Plasmoid-mediated reconnection in relativistic plasmas can accelerate particles to non-thermal energies (e.g., \citealt{Sironi_Spitkovsky,Guo_2014,Werner_2015}), responsible for the high-energy emission in many environments of compact objects (e.g., \citealt{Cerutti2015,Beloborodov17}). Recent studies of relativistic turbulence in collisionless plasmas have shown efficient particle acceleration \citep{Zhdankin17,Comisso1} and the formation of reconnecting current sheets, which are important for the process of initial particle acceleration \citep{Comisso1} both in 2D and 3D. The high-energy power-law tail of the distribution function has been shown to get steeper  quickly for smaller ratios of the turbulent component of the field to the guide field at the outer scale, $\delta B / B_0$. This observation further motivates the exploration of current sheet properties at moderate $\delta B \sim B_0$, when particle acceleration is efficient.}

The highly magnetized relativistic ($v_A\approx c$) MHD limit has been largely unexplored, and it is unclear whether dynamic alignment forms in this regime, \rev{and whether it plays an important role for the current sheet formation for situations where $\delta B \sim B_0$}. Neither any presence of dynamic alignment nor \rev{plasmoid unstable current sheets} were shown in the first relativistic ideal MHD simulations by \cite{Zrake_2012}.

%Global simulations of high-energy astrophysical systems are now capable of resolving the largest scale reconnection layers, e.g., equatorial current sheets beyond the light cylinder in pulsars magnetospheres \citep{Cerutti2015} or in the magnetosphere close to the event horizon of magnetically-arrested black hole accretion disks \citep{Parfrey_2019,Ripperda_2020}. Smaller scale current sheets can develop in local turbulence in these astrophysical systems, driven by plasma instabilities, resulting in particle acceleration (e.g., in kink-unstable jets \citep{Tchekhovskoy_2016,Davelaar_2020}).}
%\rev{In high-energy astrophysical source relativistic counterparts of the found long and stable current sheets are sources of high energy non-thermal particles responsible for non-thermal emission. While global simulations of astrophysical sources are capable of resolving only large scale current sheets (e.g. equatorial current sheets beyond light cylinder of a pulsar magnetosphere \citep{Cerutti2015} or jet-separating current sheet in magnetically-arrested accretion disks \citep{Ripperda_2020}). On the other hand, instabilities developed in astrophysical systems drive local turbulence which can
%potentially emerge smaller scale current sheets. For example, developing in relativistic jets kink instability \citep{Tchekhovskoy_2016} creates its own current sheet and accelerate particles \citep{Davelaar_2020}.} 

In this Letter we present numerical relativistic resistive MHD simulations of decaying turbulence in highly-magnetized plasma both in 2D and 3D. We demonstrate that dynamic alignment forms both in 2D and 3D. We show intermittent long-lived current sheets form naturally in the turbulence and become plasmoid-unstable.  %\BR{The main questions we address are a) Can non-relativistic turbulence results be extended to the relativistic limit applicable in black hole and neutron star magnetospheres, jets, and accretion disks and b) How well does the strong \SCH{strong turbulence means a bit different.. powerful?} turbulence limit $\delta B/B_0\approx 1$, accessible with our numerical simulations, deviate from theoretical predictions for the weak turbulence limit $\delta B \ll B_0$.}
%\rev{The main questions we address are a) how far in relativistic limit one can expand the results of non-relativistic theory and b) how significant is the deviation of the theoretical predictions for $\delta B/B_0\approx 1$ from the theoretical limit of $\delta B \ll B_0$.}
%We show \rev{that all statistical properties of the turbulence are well reproduced in the systems we study, and that} intermittent long-lived current sheets form naturally in the turbulence and become plasmoid-unstable. 

\rev{\section{Theoretical overview}\label{sec:mhdcomparison}}

\rev{The study of non-relativistic turbulence is usually done with a reduced MHD approach. This method employs a few assumptions: \rev{a uniform, strong, in comparison to the perturbation $\delta B$, guide field $B_0$} and incompressibility of the flow ($c_s \to \infty$, where $c_s$ is the sound speed). 
Under these assumptions, the only waves of interest are perpendicularly-polarized Alfv\'{e}n waves, propagating along the guide field. The reduced form of MHD equations in this limit reads \citep{Elsasser:1950}:
\begin{equation}
    \frac{\partial {\bf z^\pm}}{\partial t} \mp {\bf v}_A \cdot \nabla {\bf z^\pm} = - {\bf z^\mp} \cdot \nabla_\perp {\bf z^\pm} - \nabla P / \rho_0,~~~{\nabla\cdot\bf z^\pm} = 0,
\end{equation}
where $P$ is the total pressure, and ${\bf z^\pm}=\delta {\bf v} \pm \delta {\bf B}/\sqrt{4\pi \rho_0}$ are the Elsasser fields, representing counter-propagating Alfv\'{e}n waves.} 

\rev{The ideal relativistic MHD equations consist of mass and stress-energy conservation laws, and the induction equation for the magnetic field evolution:
\begin{equation}
    \partial_\nu(\rho u^\nu)=0,~~\partial_\nu T^{\mu \nu}=0,~~\partial_\nu(b^\mu u^\nu-b^\nu u^\mu)=0.
\end{equation}
Here $\mu,~\nu$ are 4-dimensional space-time indices, such that $u^\mu=(\Gamma,\Gamma {\bf v})$ is the four-velocity vector, $\Gamma$ is Lorentz factor,  and $T^{\mu \nu}$ is the stress-energy tensor 
\begin{equation}
    T^{\mu \nu} = \mathcal{E} u^\mu u^\nu +\left(P+\frac{b^2}{2}\right)\eta^{\mu \nu} - b^\mu b^\nu
\end{equation}
with $\mathcal{E}=\rho \omega c^2+ b^2$ and $\omega=1+({\gamma}/(\gamma-1)){P}/{\rho}$ is the relativistic enthalpy, $\eta^{\mu \nu}={\rm diag}\{-1,1,1,1\}$, the flat-spacetime Minkowski metric. $b^\mu$ is the magnetic field four-vector 
\begin{equation}
    b^\mu = \frac{1}{\sqrt{4\pi}}\left(\Gamma ({\bf v}\cdot{\bf B}),\frac{B^i}{\Gamma}+\Gamma \frac{({\bf B}\cdot{\bf v})v^i}{c^2} \right),
\end{equation}
and $b^2=b^\mu b_\mu$. 
Introducing the relativistic Elsasser fields 
\begin{equation}
    z^\mu_\pm=u^\mu\pm b^\mu/\sqrt{\mathcal{E}}
    \label{relz}
\end{equation} and modified pressure term, $\Pi=(2P+b^2)/(2\mathcal{E})$, one can rewrite the relativistic MHD equations in the Elsasser-type form: \citep{Chandran_2018,TenBarge2021}
\begin{equation}
    \partial_\nu(z_\pm^\mu z_\mp^\nu + \Pi \eta^{\mu \nu})+\left(\frac{3}{4}z_\pm^\mu z_\mp^\nu+\frac{1}{4}z_\mp^\mu z_\pm^\nu+\Pi\eta^{\mu\nu}\right)\frac{\partial_\nu \mathcal{E}}{\mathcal{E}} = 0.
\end{equation}}

\rev{In contrast to the non-relativistic case, one cannot formally introduce an incompressible limit in relativistic MHD wherein there is a maximum speed of propagation, $c$. The finite speed of light prevents easy elimination of the fast magnetosonic modes \citep{Takamoto_2017}. However, it is still possible to order them out in the highly anisotropic limit, $k_{\perp}\gg k_{||}$, \rev{which implies $\omega_F \sim kc \gg \omega_A \sim k_{||}c$}, where $\omega_F$ and $\omega_A$ are the frequencies of the fast magnetosonic and Alfv\'{e}n modes, correspondingly (\citealt{TenBarge2021}). Here, we are interested in the highly-magnetized limit, where the hot magnetization parameter, $\sigma$, is large:
\begin{equation}
\sigma = \frac{\langle B^2\rangle}{4\pi \rho c^2\omega} \gg 1.
\end{equation}
This implies $b^2\gg P$, or $\Pi=1/2$, and the elimination of the slow magnetosonic mode. The second term of the relativistic Elsasser variable $z^\mu_\pm$ is a unit vector in the direction of the four-magnetic field vector as $\mathcal{E}\approx \sqrt{b^2}$ in this limit.}

\rev{\cite{TenBarge2021} discusses the Elsasser-type equations of the relativistic MHD in the highly anisotropic limit, $k_{\perp}\gg k_{\parallel}$, constructed in the average fluid rest frame $\langle u^i \rangle=0$ \citep{Chandran_2018}. In three-vector form the result is particularly similar to the reduced non-relativistic MHD equations:
\begin{equation}
    \frac{\partial \delta {\bf z}_\pm}{\partial t} \mp {\bf v}_A \cdot \nabla \delta {\bf z}_\pm = -\delta {\bf z}_\mp \cdot \nabla_\perp \delta {\bf z}_\pm - \nabla_\perp \delta \Pi.
    \label{rel3form}
\end{equation}
\begin{equation}
    \delta {\bf z}_\pm =\delta {\bf v} \pm \frac{\delta {\bf B}_\perp}{\sqrt{\mathcal{E}_0}},~\delta \Pi = -\frac{2\delta P+\delta \rho}{2 \mathcal{E}_0}+\frac{2 P_0+\rho_0}{2 \mathcal{E}_0}\frac{\delta \mathcal{E}}{\mathcal{E}_0}. \nonumber
\end{equation}
\nrev{Due to the close resemblance between the Newtonian and relativistic set of reduced MHD equations, once the anisotropic cascade reaches sufficiently small scales, where $k_{\perp}\gg k_{\parallel}$ is satisfied, one can expect that  relativistic MHD turbulence is statistically similar to its Newtonian counterpart.}
In the case of interest, $\sigma\gg1$, the last term of (\ref{rel3form}) is negligible, and the Alfv\'{e}n speed is close to the speed of light, $v_A\approx c$, such that the equations are particularly simple: 
\begin{equation}
    \frac{\partial \delta {\bf z}_\pm}{\partial t} \mp \nabla_{||} \delta {\bf z}_\pm = - \delta {\bf z}_\mp \cdot \nabla_\perp \delta {\bf z}_\pm.
    \label{zpzmsimplified}
\end{equation}
\nrev{The applicability of equation (\ref{zpzmsimplified}) is limited to regimes where $\delta P \ll \delta B^2$. However, when plasma is heated to relativistic temperatures, e.g., in reconnection layers, this assumption is not justified, at least locally.} Since there is no a formally incompressible limit in relativistic systems, and our interest in systems with $\delta{B} \sim B_0$, which is particularly challenging to explore analytically, we turn to numerical simulations to confirm these expectations.}

\rev{An important feature of the highly magnetized MHD turbulence is the overall dominance of the magnetic \nrev{and electric} field fluctuations, $\delta E_B$, over the kinetic energy, $\delta E_{\rm kin}$. This can be seen from the following relations for a single Alfv\'{e}n wave in the relativistic regime ($\sigma\gg 1, v_A\sim c$):
\begin{eqnarray}
    \delta E \sim \frac{\delta  v}{c}~B_0 \sim \frac{v_A}{c}~\delta  B \sim \delta  B \Rightarrow \delta v \sim c \frac{\delta B}{ B_0} < c \nonumber \\
    \delta E_{\rm kin}\sim\rho \delta v^2 \sim \delta B^2 \frac{\rho c^2}{B_0^2} \sim \delta B^2 \frac{1}{\sigma_{\rm {cold}}} \ll  \delta E_B, \nonumber
\end{eqnarray}
where the inequality is used,
\begin{equation}
\sigma_{\rm {cold}}= \frac{B^2}{4\pi \rho_0 c^2} >  \frac{B^2}{4\pi \rho_0 c^2 \omega} = \sigma.
\end{equation}}

\rev{Current sheets are important dissipative structures in magnetized turbulence, and it is useful to compare their behavior in non-relativistic and relativistic regimes. In a near-stationary current sheet, the reconnection rate is the ratio of the inflow velocity to the outflow velocity $v_{\rm in}/v_{\rm out}$ (\citealt{parker_1957}; \citealt{sweet_1958}). If the plasma can be
assumed incompressible, it then follows that $v_{\rm in}/v_{\rm out}=\delta/L$ where $\delta$ is the thickness and $L$ is the length of the current sheet. The thickness of a Sweet-Parker current sheet is determined from continuity of the resistive and ideal electric fields as $\delta=\eta/v_{\rm in}$, where $\eta$ is the resistivity. The outflow speed can be approximated as the Alfv\'{e}n speed $v_{\rm out} \sim v_{\rm A}$,  which in a relativistic plasma is $v_{\rm A} \sim c$. This results in a relativistic reconnection rate of $v_{\rm in}/v_{\rm out} \sim v_{\rm in} / c \sim \eta/(L v_{\rm in})$, i.e., a result identical to the non-relativistic case (\citealt{Lyubarsky2005}). The reconnection rate in a Sweet-Parker sheet then scales as $\sim S^{-1/2}$, where $S=L v_{\rm A} / \eta$ is the Lundquist number. \rev{For large Lundquist numbers, typical in astrophysical sources,} reconnection is mediated by the plasmoid instability, which in non-relativistic settings gets triggered at $S_{\rm crit}\geq 10^{4}$, leading to a saturation of the reconnection rate at $\approx 0.01$ \citep{Loureiro2007, bhattacharjee2009}. It was shown semi-analtyically and numerically, by solving the full set of resistive relativistic MHD equations, that this result holds for highly magnetized relativistic plasmas \citep{DelZanna,Ripperda:2019a}.}

\rev{Since magnetic field fluctuations dominate over kinetic energy, resistive dissipation dominates over viscous dissipation in the inertial range of turbulence. We show that resistive dissipation in highly-magnetized MHD plasmas is also dominant in the exhausts of reconnection layers. One can see this by comparing the rate of resistive energy dissipation $\sim  \eta B^2 / 4\pi\delta_{\rm exhaust}^2$, to the rate of viscous dissipation $\sim 2 \nu E_{\rm kin} / \delta_{\rm exhaust}^2$, where $\nu$ is the viscosity, $E_{\rm kin}$ is the kinetic energy in the exhaust region, $B^2/8 \pi$ is magnetic energy and $\delta_{\rm exhaust}$ is the typical width scale in the  exhaust region. The ratio of resistive to viscous dissipation rates is then $\eta / \nu \cdot (B^2 / E_{\rm kin}) \sim (B^2 / E_{\rm kin})$\footnote{\rev{Our argument is applicable for the case of a scalar viscosity $\nu$, i.e., independent of the gas pressure which is high in the exhaust region. Whether this assumption is accurate for the effective viscosity of a collisionless relativistically hot plasma is an interesting question for further investigation.}}. In a non-relativistic plasma, this ratio is always of order $\mathcal{O}(1)$ since $E_{\rm kin} \sim \rho v_{\rm A}^2$ and $v_{\rm A}^2 \sim B^2/\rho$. However, in highly magnetized relativistic plasma, $E_{\rm kin} \sim \rho \gamma_{\rm exhaust} c^2 \sim \rho \sqrt{\sigma} c^2$ \nrev{\citep{Lyubarsky2005}} while $B^2\sim \rho \sigma c^2$, and hence the ratio between resistive and viscous dissipation rates is proportional to $\sqrt{\sigma}\gg 1$ such that resistive dissipation dominates.}

\section{Numerical Method and Setup} \label{sec:method}

We solve the set of special relativistic resistive MHD (SRRMHD) equations with the Black Hole Accretion Code ({\tt{BHAC}}, \citealt{BHAC1, BHAC2}) and an Implicit-Explicit (IMEX) time stepping scheme to evolve the stiff resistive Ohm's law \citep{Ripperda:2019a,resBHAC}. We employ a constant and uniform resistivity $\eta$, which provides the simplest prescription to allow resolved magnetic reconnection.

 The SRRMHD equations are numerically evolved in a periodic domain of size $L^2$ in 2D and $L^2\times L_z$ in 3D. We initialize an out-of-plane (in the $\bf{\hat{z}}$-direction) guide magnetic field and an in-plane ($x-y$) magnetic field perturbation $\delta {\bf B}_\perp$:
\begin{eqnarray}
\delta B_x&=&\sum_{m=1}^N \sum_{n=1}^N \beta_{m n} n \sin(k_m x + \phi_{mn}) \cos(k_n y + \varphi_{mn}),\nonumber\\
\delta B_y&=&-\sum_{m=1}^N \sum_{n=1}^N\beta_{m n} m \cos(k_m x + \phi_{mn}) \sin(k_n y + \varphi_{mn}),\nonumber
\end{eqnarray}
where $\beta_{m n}=2 \delta B_\perp/(N\sqrt{m^2+n^2})$, $k_m=2\pi m/L$, and $\phi_{mn},~\varphi_{mn}$ are random phases. We set $N=8$ initial waves in each direction for 2D runs (64 initial modes) and $N=4$ for 3D runs, in order to allow for a larger inertial range in 3D simulations. \rev{The  outer (or energy containing) scale is then $l_\perp=L_0/8$ for 2D simulations and $l_\perp=L_0/4$ for 3D. The turbulence at smaller scales forms self-consistently via energy cascading.} In 3D we modulate $\delta {\bf B}_\perp$ with two modes $\propto \sin(k_l z + \psi_{mnl})$, where $\psi_{mnl}$ is also a random phase. The normalization coefficient is then $\beta_{m n l}=2\sqrt{2} \delta B /(N\sqrt{N_z}\sqrt{m^2+n^2})$. We initialize the plasma at rest, with velocity field ${\bf v}=0$, and with a uniform gas pressure $p_0$ and rest mass density $\rho_0$. \rev{We set an adiabatic index $\gamma=4/3$, assuming an ideal relativistic gas.} Similar initial conditions \rev{for the magnetic field} were employed in relativistic particle-in-cell (PIC) \citep{Comisso1,Nattila} turbulence simulations. For all simulations we set $L=1$.

In order to characterize the strength of both the guide and the in-plane magnetic field we introduce two magnetization parameters
\begin{equation}
\sigma_0=\frac{\langle B_z^2 \rangle}{4\pi \rho_0 c^2\omega},~~~\delta \sigma = \frac{\langle \delta B_\perp^2\rangle}{4\pi \rho_0 c^2\omega},
\end{equation}
%\crossout{where $\omega=1+({\gamma}/(\gamma-1)){P_0}/{\rho_0}$ is the relativistic enthalpy, and angular brackets denote spatial average}. 
A summary of the performed runs is given in Table \ref{tab:table1}. We employ an elongated box with $L=1, L_z=3$ for run 3D[d] to enforce the critical balance condition  \mbox{$\delta B_\perp/L\approx B_0/L_z$} at the outer scale.

\begin{table}[b]
\caption{\label{tab:table1}Summary of simulation parameters. }
\begin{tabular}{ccccccc}
\hline \hline
\textrm{Sim}&
\textrm{Res}&
\textrm{$\eta$}&
\textrm{$\sigma_0$}& 
\textrm{$\delta\sigma$}&
\textrm{$L, L_z$}&
\textrm{Grid}\\
\hline
 2D[a] & $65536^2$ & $10^{-6}$ & 5 & 5 & $1^2$ & \textrm{AMR}\\
 2D[b] & $65536^2$ & $10^{-5}$ & 5 & 5 & $1^2$ & \textrm{AMR}\\
 2D[c] & $32768^2$ & $10^{-6}$ & 5 & 5 & $1^2$ & \textrm{AMR}\\
 2D[d] & $32768^2$ & $10^{-6}$ & 1 & 1 & $1^2$ & \textrm{AMR}\\
 2D[e] & $32768^2$ & $10^{-6}$ & 1 & 5 & $1^2$ & \textrm{AMR}\\
 2D[f] & $65536^2$ & $10^{-6}$ & 5 & 5 & $1^2$ & \textrm{Uni}\\
 \hline
 3D[a] & $3200^3$ & $10^{-6}$ & 1 & 5 & $1^2 \times 1$ & \textrm{Uni}\\
 3D[b] & $2048^3$ & $10^{-6}$ & 5 & 5 & $1^2 \times 1$ & \textrm{Uni}\\
 3D[c] & $2048^3$ & $10^{-6}$ & 1 & 1 & $1^2 \times 1$ & \textrm{Uni}\\
 3D[d] & $2048^3$ & $10^{-6}$ & 9 & 1 & $1^2 \times 3$ & \textrm{Uni}\\
 \hline \hline
\end{tabular}
\end{table}

We set the resistivity to either $\eta=10^{-5},~10^{-6}$ in the 2D setup, which corresponds to Lundquist numbers $S \approx 10^{4}, 10^{5}$ for the largest current sheets of length $L_{\rm cs}\approx0.1$, and $v_A/c\approx 1$. The simulation with $\eta=10^{-6}$ is well-above the critical Lundquist number limit $S_{\rm crit}$, while the simulation with $\eta=10^{-5}$ is approximately at the limit $S\approx S_{\rm crit}$. \rev{Potentially, current sheets can become plasmoid unstable at a smaller critical Lundquist number in a turbulent flow \citep{Loureiro_2009}. We explore whether this effect is significant in 2D relativistic turbulence with our $\eta=10^{-5}$ simulation.}
 
In order to ensure that the resistive length scales are resolved, and results are converged with numerical resolution, we develop a novel adaptive mesh refinement (AMR) strategy (see Appendix \ref{Appendix:AMR}). We benchmark our 2D results with a short simulation (until $t=0.5L/c$) on a uniform grid, with a resolution of $65536^2$. In 3D it is impossible to fully converge due to numerical limitations, and instead we employ the highest feasible resolution that allows to capture the development of the plasmoid instability in the longest current sheets. One high-resolution run is performed with $3200^3$ grid points to probe the formation of plasmoid chains. We additionally present a study with different values of the magnetization parameter at a resolution of $2048^3$ grid points.

The SRRMHD algorithm relies on viscosity $\nu$ at the grid level, such that the magnetic Prandtl number ${\rm Pr_m}=\nu/\eta \ll 1$ for 2D simulations, and ${\rm Pr_m} \lesssim 1$ for 3D simulations with a marginally resolved resistive scale, assuming resistive and viscous scales are similar \rev{and governed by the finite grid resolution}. \rev{This choice is further motivated by the fact that viscous effects are subdominant for highly-magnetized plasmas, as we have demonstrated in Section \ref{sec:mhdcomparison}.} 
%\crossout{In the highly magnetized, $\sigma\gg1$, regime we expect most of the dissipation to be governed by resistive effects rather than by viscosity, because the energy of magnetic field fluctuations dominates over kinetic energy in high-$\sigma$ turbulence. This can be seen from the following relations for a single Alfv\'{e}n wave in the relativistic regime ($v_A=c$):
%\begin{eqnarray}
%    \delta E \sim \frac{\delta  v}{c}~B_0 \sim \frac{v_A}{c}~\delta  B \Rightarrow \delta v \sim c \frac{\delta B}{ B_0} < c \nonumber \\
%    \rho \delta v^2 \sim \delta B^2 \frac{\rho c^2}{B_0^2} \sim \delta B^2 \frac{1}{\sigma_{\rm {cold}}} \ll \delta  B^2, \nonumber
%\end{eqnarray}
%where the inequality is used,
%\begin{equation}
%\sigma_{\rm {cold}}= \frac{B^2}{4\pi %\rho_0 c^2} >  \frac{B^2}{4\pi \rho_0 c^2 %\omega} = \sigma.
%\end{equation}
%}

\section{Results} \label{sec:results}

We present the results of simulations which we run until $t=2L/c$, such that the turbulence is fully developed and settles to a quasi-steady state. For the case of decaying turbulence considered in this Letter, we define a quasi-steady state when the spectral slope is constant for at least one outer scale eddy turnover time, $\sim L/c$, while the total energy \mbox{$E_B=\int {\bf B(x)}^2\nrev{/8\pi} d{\bf x}=\int {\bf B_k(k_\perp)}\cdot{\bf B^*_k(k_\perp)} \nrev{/8\pi} d{\bf k}$} dissipates. Here, ${\bf B_k}$ is the amplitude of the Fourier mode of the magnetic field with wavenumber $k$. It takes  $\Delta t\approx0.3L/c$ for the energy to cascade from the initial low $k$ modes to the resistive scale. At $\Delta t\approx 1-2L/c$ the spectrum flattens until it reaches a quasi-steady state. This behavior of the power spectrum is illustrated in the attached \href{https://youtu.be/n7SZigrJ9kk}{video}\footnote{Direct link: \url{https://youtu.be/n7SZigrJ9kk} }. In order to test convergence of the simulation, we compare spectra of magnetic energy, \mbox{$E(k)dk=\sum_{{\bf k} \in dk}{\bf B}_{\bf k} \cdot {\bf B}_{\bf k}^*\nrev{/8\pi}$}, for different resolutions: 
%if the \crossout{highest wave-numbers, $k_\perp$, of the inertial range coincide,}
\rev{if the onset of the inertial range cutoff  does not change with increasing resolution (the vertical lines in Figure~\ref{fig:AMR}c), i.e., if the cutoff is determined by the resolved resistive scale,} the simulation is considered converged. More details about the AMR strategy and convergence tests are presented in the Appendix \ref{Appendix:AMR}.

\begin{figure*}
\includegraphics[width=1.02\linewidth]{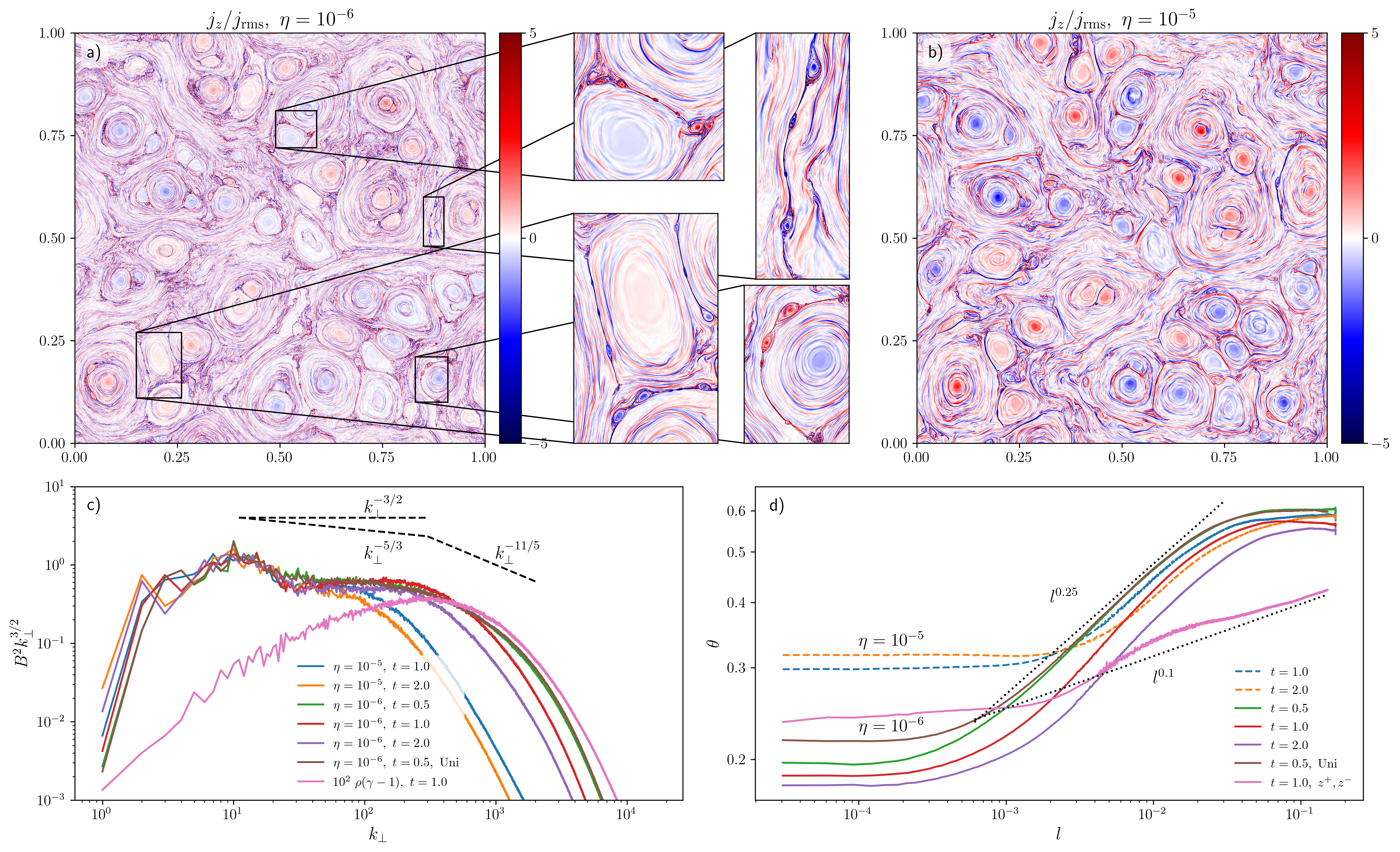}
\caption{2D SRRMHD runs of highly-magnetized decaying turbulence.  The top row shows snapshots of the out-of-plane normalized electric current at $t=1$ for a) simulation 2D[a], with the resistivity value $\eta=10^{-6}$ corresponding to the typical Lundquist number $S\approx10^5$ for the longest current sheets; b) simulation 2D[b], $\eta=10^{-5}$, $S\approx10^4$. Insets show zooms into the snapshot of simulation 2D[a], highlighting plasmoid-unstable current sheets. The bottom row shows statistical properties of the 2D turbulence: c) the spectrum of the \rev{normalized} magnetic \rev{ and kinetic (multiplied by 100) energies} and d) the dynamic alignment \rev{angle} at different times during the simulation $t=0.5,t=1.0,t=2.0$, in  simulations 2D[a,b] \rev{and alignment angle for Elsasser field, $\theta^{z^+,z^-}$, for 2D[a] at $t=2.0$}. The results of a uniform grid simulation 2D[f] at $t=0.5$ are presented to show numerical convergence of the AMR criteria.}
\label{fig:2d}
\end{figure*}

In Figure \ref{fig:2d}a we present the distribution of the out-of-plane electric current density $j_z\sim (\nabla\times{\bf B})_z$ for a 2D simulation with \rev{an effective resolution of} $65536^2$ grid points, 2D[a], and 3 AMR levels, for a resistivity $\eta=10^{-6}$. Here, we set magnetizations $\sigma_0=\delta \sigma=5$, equivalent to a total magnetization $\sigma=10$. 
Very long current sheets emerge at the interfaces of large merging eddies. The length of a current sheet is mainly defined by the size of the largest eddies present in the system. Estimating the length of these current sheets as $L_{\rm sheet}/L \approx 0.1$ and accounting for the relativistic Alfv\'{e}n speed $v_A/c=\sqrt{\sigma/(\sigma+1)}\approx 1$, we find the Lundquist number to be $S\approx 10^5 \gg S_{\rm crit}=10^4$. These current sheets are  plasmoid-unstable and break up into current sheets of smaller length scales, such that their Lundquist numbers $S_{\rm local} \approx  S_{\rm crit}$. This results in a maximum number of $\sim 10$ plasmoids, which is consistent with results shown in Figure \ref{fig:2d}a. We also perform simulations for $\sigma_0=\delta \sigma=1$ and $\sigma_0=1, \delta \sigma=5$ and resistivity $\eta=10^{-6}$. Plasmoid-unstable current sheets form ubiquitously for all of these settings. \rev{For resistivity $\eta=10^{-5}$ we observe only few plasmoids (for the longest current sheets) in the whole domain indicating that the critical Lundquist number $S_{\rm crit}\approx10^4$ holds for the plasmoid instability in current sheets in a 2D turbulent flow.}  By varying numerical resolution, we find that the onset of the plasmoid instability occurs at lower resolution for the cases with weaker guide field, $\sigma_0 \leq \delta \sigma$, motivating our choice to perform our highest resolution 3D simulation for $\sigma_0/\delta \sigma=1/5$ (run 3D[a]). 

2D and 3D \rev{weak guide field (3D[a])} simulations show pronounced reconnection-mediated mergers of smaller eddies. This process has also been recently observed in simulations of merging non-helical flux tubes \citep{Zhou_2020}. Our very long 2D simulations with a (smaller) resolution of $3200^2$ demonstrate that the terminal state of the turbulence has two large eddies of opposite magnetic helicity $\int {\bf A}\cdot  {\bf B} d{\bf x}$ remaining in the simulation box. 3D simulations show similar behavior. In order to identify  current sheets, we choose a threshold in the current density, $\xi$, and consider a point ${\bf x}$ to be in the current sheet if $j_z({\bf x})>\xi j_{\rm rms}$, where $j_{\rm rms}$ is the root-mean-square of the electric current $j_z$ in the domain (\citealt{Zhdankin}). The long current sheets have an intermittent nature and occupy about 0.2-0.5$\%$  of the domain in 2D, yet they are responsible for $20-25\%$ of the magnetic field dissipation, $\propto \eta j^2$, for $\eta=10^{-6}$. Our results are insensitive to the exact value of $\xi$, as long as $\xi \gtrsim 5$. For a larger value of resistivity, $\eta=10^{-5}$, only few plasmoids form in the whole simulation box (see Figure \ref{fig:2d}b). Comparing to the $\eta=10^{-6}$ case, the current sheets for $\eta=10^{-5}$ are thicker and, hence, have a lower current density amplitude. In this case we find only $10\%$ of the dissipation to happen in the localized current sheets. 

The anisotropic properties of the turbulence can be quantified by measuring the dynamic alignment \rev{angle} of eddies in the plane perpendicular to the guide field \citep{Boldyrev_2005, Boldyrev_2006}. We employ a Monte-Carlo method to compute dynamic alignment angle as a function of a point-separating vector. Following the method proposed by \cite{Mason_2006, Perez_2012}, we compute two structure functions \mbox{$S^{\nrev{1}}_1(l)=\langle \delta {\bf v}_\perp({\bf l}) \times {\bf B}_\perp({\bf l})\rangle$}, \mbox{$S^{\nrev{1}}_2(l)=\langle |\delta {\bf v}_\perp({\bf l})| |{\bf B}_\perp({\bf l})|\rangle$}, where $\delta {\bf v}_\perp ({\bf l})$ and   $\delta {\bf B}_\perp ({\bf l})$ are increments of the velocity and the magnetic field perpendicular to the \textit{local} guide field  at scale $l$. The alignment angle is defined as
\begin{equation}
    \theta(l) \rev{\equiv \theta^{v,B}(l)} \equiv \frac{l}{\xi} =\frac{S^{\nrev{1}}_1}{S^{\nrev{1}}_2},
\end{equation}
where $l$ and $\xi$ are sizes of the eddy in the guide field-perpendicular plane. \nrev{Note that both $S^1_1$ and $S^1_2$ are first order structure functions $S^n(l)=\langle |f({\bf r+l})-f({\bf r})|^n \rangle$, and one can define the alignment angle for any $n$ as \mbox{$\theta\approx\left(S^n_1(l)/S^n_2(l)\right)^{1/n}$}. It turns out that the slope of the alignment angle function is dependent on the order of the structure function $n$ \citep{Mallet2016} revealing the intermittent nature of dynamic alignment \citep{frisch_1995}. We present more details on the intermittency of dynamic alignment in Appendix \ref{Appendix:Int}.}

%\rev{Some of the theories of the alignment formation employ the mutual shear of counter-propagating Elsasser fields $\delta {\bf z}_\pm$ (see, e.g. \citep{Chandran_2015}). It is predicted that these two fields also create progressively decreasing alignment angle, but the slope is flatter. 

The slope of the dynamic alignment \rev{angle as a function of the size of eddies} is tightly connected to the power-law of the magnetic energy spectrum, $P_{\bf B}(k)$, which is predicted to be $k_\perp^{-3/2}$ for turbulence that is anisotropic in the plane perpendicular to the guide field \citep{Boldyrev_2006}. By introducing a non-linear time, $\tau_c$ \citep{Boldyrev_2005}, we can relate the power spectrum with the alignment angle:
\begin{flalign}
&\begin{cases}
\frac{\delta B_l^2}{\tau_{c}}\sim\varepsilon\\
\tau_{c}\sim\frac{l}{\delta B_l \sin{\theta}},
\label{SpectAlign}
\end{cases}&\\
&E(k_\perp)\sim\delta B_l^2 k_\perp^{-1}\sim \varepsilon^{2/3}k_\perp^{-5/3} \sin{\theta}^{-2/3},&\nonumber
\end{flalign}
where $\varepsilon$ is the energy cascading rate, $\delta B_l$ is the increment of the magnetic field at a scale $l$ in the plane perpendicular to the guide field. For $\sin{\theta(l)}\sim l^{1/4} \sim k_\perp^{-1/4}$, it reproduces Boldyrev's spectrum \mbox{$E(k_\perp)\sim k_\perp^{-3/2}$}. In the case of no alignment being present, $\theta(l)\sim {\rm const}$, it reduces to the \cite{GS95} spectrum, $E(k_\perp)\sim k_\perp^{-5/3}$.

\begin{figure*}
\includegraphics[width=1.0\linewidth]{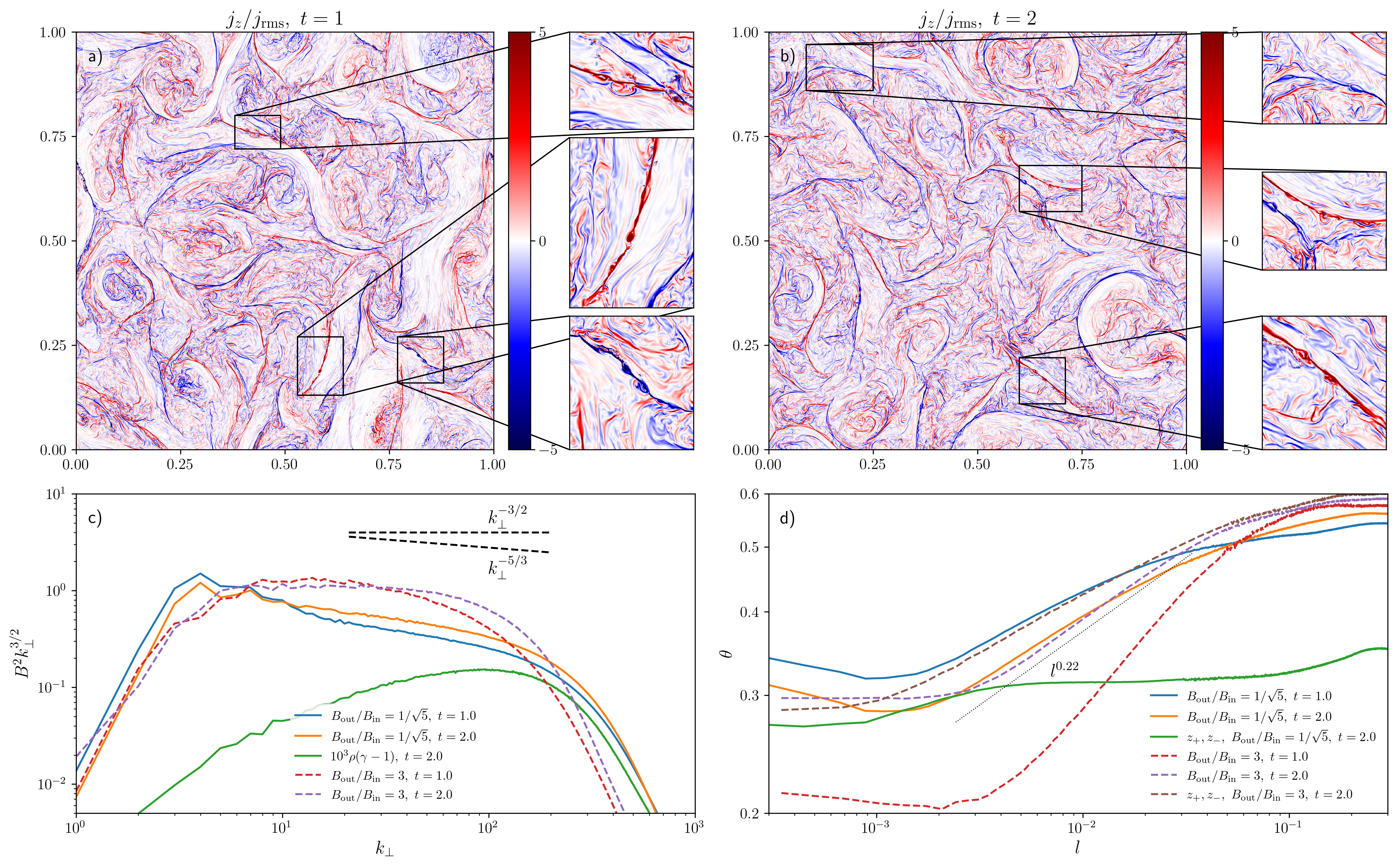}
\caption{3D SRRMHD runs of highly-magnetized decaying turbulence. The top row shows snapshots of the out-of-plane normalized electric current $j_z$ for run 3D[a] at a) $t=1.0$ and b) $t=2.0$. Insets of both figures show zooms into  plasmoid-unstable current sheets. The bottom row shows statistical properties of the 3D turbulence: c) the spectrum of \rev{normalized} magnetic \rev{ and kinetic (multiplied by $10^3$) energies} and d) the dynamic alignment \rev{angles $\theta$} for runs 3D[a] (solid lines) and 3D[d] (dashed lines) at two different times $t=1.0,~t=2.0$, \rev{and alignment angles for the Elsasser fields, $\theta^{z^+,z^-}$, at $t=2$ for runs 3D[a], 3D[d]}.}
\label{fig:alignment3d}
\end{figure*}

In Figure \ref{fig:2d}c we show the magnetic power spectrum \rev{in the steady state} and multiply the result by $k_\perp^{3/2}$, to make the difference between the  power law indices $-3/2$ and $-5/3$ more pronounced. Figure \ref{fig:2d}c clearly demonstrates that the spectrum is closer to $k_\perp^{-3/2}$ in the inertial range. 
%\crossout{ Although a steady state cannot be correctly defined in decaying turbulence, the slope of the spectrum is almost constant at $t \geq 0.5 L/c$, which justifies the comparison with the steady state theory.}
\rev{We define the steady state of decaying turbulence when the spectral slope is constant in time, at $t \geq 0.5 L/c$, allowing us to compare our results with steady state theory. Note that the total energy $\epsilon_{\rm B}(t) = \int (B^2-B|_{{\bf k}=0}^2)\nrev{/8\pi} dV$ decreases in time while the normalized spectrum $\Tilde{E}(k)dk=\sum_{{\bf k} \in dk}{\bf B}_{\bf k} \cdot {\bf B}_{\bf k}^*/\nrev{8\pi}\epsilon_{\rm b}$, \rev{which we present in all spectrum plots}, is constant in time.} It is worth mentioning that in non-relativistic reduced MHD simulations one typically analyzes the spectrum of the total kinetic and magnetic energy \citep{Perez_2012}. In our highly magnetized relativistic simulations however, %\crossout{$\rho{\bf v}^2_{\bf k}(k_\perp)\ll{\bf B}^2_{\bf k}(k_\perp)$}
\rev{$\rho(\gamma-1)_{\bf k}(k_\perp)\ll{\bf B}^2_{\bf k}(k_\perp)$}, and we confirmed that the contribution of the kinetic energy is negligible \rev{for both 2D and 3D simulations (see spectra in Figures \ref{fig:2d}c and \ref{fig:alignment3d}c)}. \rev{To preserve the kinetic to magnetic field energy ratio, we also normalize the kinetic energy by $\epsilon_{\rm b}(t)$.} In agreement with the $k_\perp^{-3/2}$ power spectrum, the \nrev{${\bf v}-{\bf B}$} dynamic alignment (Figure \ref{fig:2d}d) demonstrates a perfect match with \nrev{Boldyrev's} prediction, $\theta(l)\sim l^{1/4}$, at the intermediate scales, \mbox{$l_{\rm res}\lesssim l \lesssim  l_{\rm max}$}, where $l_{\rm max}=L_0/8$ is defined by the number of modes in the initial conditions, and $l_{\rm res}$ is defined by the resistive scale. 

\rev{\cite{Chandran_2015} proposed that mutual shear of counter-propagating Elsasser fields $\delta {\bf z}_\pm$ is responsible for the dynamic alignment. They predict that these two fields create a progressively decreasing alignment angle, while the slope becomes flatter. To test this
hypothesis, we measure the alignment angle between two Elsasser fields:
\begin{equation}
    \theta^{z^+,z^-}=\frac{\langle\delta {\bf z_\perp^+}\times{\delta \bf z_\perp^-}\rangle}{\langle|\delta {\bf z_\perp^+}||\delta {\bf z_\perp^-}|\rangle}.
\end{equation}
Straightforward application of the non-relativistic Elsasser field expression, $\delta {\bf z}_{\pm}=\delta {\bf v} \pm \delta {\bf B}/\sqrt{4\pi \rho}$, results in $\delta {\bf z_\perp^+}\times{\delta \bf z_\perp^-} \sim \delta {\bf v} \times \delta {\bf B}$, while $|\delta {\bf z_\perp^+}||\delta {\bf z_\perp^-}|\sim |\delta B|^2$, giving that their ratio $\theta^{z^+,z^-} \sim \delta v/\delta B \ll 1$ in highly magnetized plasma. However, one should use the relativistic formulation of Elsasser fields (\ref{relz}) in this regime, where ${\bf u}$ and ${\bf b}/\sqrt{\mathcal{E}}$ can be comparable. %Note that the increments $\delta {\bf z_\perp^{\pm}}$ are measured in the plane perpendicular to the guide three-field ${\bf B}_0$ (while the magnetic field four-vector is used in ${\bf z}^{\pm}$). 
The dynamic alignment angle between the relativistic Elsasser fields is flatter than $l^{0.25}$ at $t=2$, for $\eta=10^{-6}$ (Figure~\ref{fig:2d}d). The average slope is close to the $l^{0.1}$ result, as predicted by \cite{Chandran_2015}, although it displays an unexpected break at intermediate scales.}
% \nrev{(also, their theory predicts the slope of the ${\bf v}-{\bf B}$ dynamic alignment angle to be $l^{0.21}$ )}.}

The smallest averaged dynamic alignment \rev{angle}, $\theta^{\nrev{{\bf v}-{\bf B}}}$, in the simulation with $\eta=10^{-6}$ is $0.175$, and it is approximately constant for small %\crossout{eddies}
\rev{scales}. \rev{Deviations from \nrev{Boldyrev's} scaling $l^{0.25}$ are visible at scales where resistive effects become important. Note that this is also where the inertial range of the spectra ends.} The plasmoid-unstable current sheets we observe in the simulation 
%\crossout{require} 
\rev{possess} much smaller alignment angles $\theta\approx 0.01$,
%\crossout{to form}, 
%according to \cite{Loureiro2007}. 
\rev{in accordance with \cite{Loureiro2007}}. The presence of such current sheets with alignment angles of an order of magnitude smaller than the minimal averaged alignment angle that we find, implies the intermittent nature of these sheets \citep{Dong2018}.
Formation of intermittent plasmoid-unstable current sheets can be responsible for a steepening of the spectrum at the end of the inertial range,  which we observe in the range $k_\perp\approx300-1200$ at $t=1$ in Figure \ref{fig:2d}c. However, we assume that the scale separation in our simulations is not enough to robustly confirm the $k_\perp^{-11/5}$ prediction by \cite{Boldyrev_Loureiro} and \cite{Mallet_2017} for the non-relativistic reconnection-mediated regime. We also do not observe the increase of the alignment \rev{angle} at small scales $l$ corresponding to wave-vectors $k_\perp$ in the steepening range, as predicted in \cite{Boldyrev_Loureiro}. 

Since the onset of the plasmoid instability occurs at lower resolution in 2D simulations if a weaker guide field is assumed, we run a 3D simulation (3D[a]) with $\sigma_0=1$, $\delta \sigma=5$, and highest resolution of $3200^3$ grid points. For 2D simulations we confirm that full plasmoid chains form for smaller values of $\delta B_\perp/B_0$ as well, but higher resolutions are required to resolve the instability. \rev{We refer to the case with initial $\delta B_\perp/B_0 = \sqrt{5}/1$ (run 3D[a]) as a weak guide field, and to the case with initial $\delta B_\perp/B_0 = 1/3$ (run 3D[d]) as a strong guide field. We note that by $t=1-2$, when we analyze the simulations, the turbulent component of the field decayed to $\delta B_\perp/B_0 \lesssim 1$ (3D[a]) and $\delta B_\perp/B_0 \sim 0.2$ (3D[d]).}  %\crossout{Since the energy cascade is developing mostly in $k_{\perp}$, we reduce the full 3D analysis to a 2D analysis in the $x-y$ plane.} 

\rev{For the strong guide field case, the energy cascade is developing mainly in $k_{\perp}\perp \hat{\bf z}$, and the full 3D analysis can be reduced to a 2D analysis in a set of $x-y$ planes \citep[e.g.,][]{Perez_2012}. For simplicity, in the case of the weak guide field we also compute the spectrum for wavevectors $k_{\perp}$ perpendicular to $\bf{B}_0$ using the same method (a more accurate calculation would use structure functions which take into account a locally varying guide field, \citealt{Cho_Vishniac}).}
%, $\langle {\bf B}_z\rangle = B_0$ due to magnetic flux conservation, and $\langle k_{\perp}^{\rm local}\rangle \perp \hat{\bf z}$.\SCH{small teaser lol} This assumption could be justified a posteriori via computation of the structure function \citep{Cho_Vishniac}. } 
In order to provide a statistically significant result, we average the 2D spectrum and dynamic alignment \rev{angle} in the set of $x-y$ planes taken at various $z$. We confirm that the spectrum and the alignment \rev{angles} are independent of the choice of the sampling planes if $N_{\rm planes} \gtrsim N_z/3$, where $N_z$ is the number of grid points in the direction along $z$. 
%In Figure \ref{fig:alignment3d} we show the dynamic alignment \rev{angle} and spectrum of 3D turbulence to be similar to the 2D results. 

We consider the turbulence at $t=2$ to be in a steady state, i.e., the dynamic alignment is fully \rev{formed} (see Figure \ref{fig:alignment3d}).
We confirm \rev{the steady state shape} of the dynamic alignment \rev{angle function} beyond $t=2$ with longer simulations at a lower numerical resolution, $2048^3$ (runs 3D[b], 3D[c]). 
%These simulations demonstrate that it takes $t\approx1.5-2$ light crossing times for the alignment and turbulence spectrum to reach a quasi steady state. 
The slope of the \rev{${\bf v}-{\bf B}$} alignment \rev{angle} is close to the predicted $l^{0.25}$ for the smaller eddies and is less pronounced for eddies of the system size scale (see Figure \ref{fig:alignment3d}d). In simulation 3D[a] with the initially weaker guide field $\delta B_\perp/B_0 = \sqrt{5}/1$, at $t\approx 1$, the alignment \rev{angle} curve is significantly shallower, consistent with the steady state in driven non-relativistic turbulence at $\delta B_\perp/B \sim 1$ \citep{Mason_2006}. At this time, the strength of turbulent fluctuations is similar to the strength of the guide field, $\langle | \delta {\bf B}_\perp | \rangle \approx \langle | {\bf B}_z | \rangle $. Further dissipation of the magnetic energy leads to $\langle | \delta {\bf B}_\perp | \rangle \approx 0.7 \langle | {\bf B}_z | \rangle $ at $t=2$, and a steeper alignment \rev{angle} curve. The spectrum of the turbulence develops simultaneously with the dynamic alignment. 
%The characteristic time scales of the formation of the spectrum in 3D are similar to ones in 2D. %Similarly to the dynamic alignment \rev{angle}, the slope of the spectrum in 3D[d] at $t=2$ is consistent with $k_\perp^{-3/2}$, while it is slightly steeper in 3D[a] with the initially weaker guide field (see Figure \ref{fig:alignment3d}c).

\rev{The slope of the $z_+-z_-$ dynamic alignment angle, $\theta^{z^+,z^-}(l)$, is comparable to $\theta(l)$ for the strong guide field (run 3D[d], $t=2$, $\langle | \delta {\bf B}_\perp | \rangle /\langle | {\bf B}_z | \rangle \approx 0.2$). For the weak guide field (3D[a], $t=2$, $\langle | \delta {\bf B}_\perp | \rangle /\langle | {\bf B}_z | \rangle \approx 0.7$), the $z_+-z_-$ alignment is very weakly pronounced. At the same time, the slope of the energy spectrum of 3D[a] is closer to $-5/3$ as predicted by Goldreich-Sridhar theory with no dynamic alignment. It could be considered as an indication that the dynamic alignment of Elsasser fields $\delta {\bf z}_+,~\delta {\bf z}_-$ rather than the one of ${\bf v}, {\bf B}$ reduces the non-linearity.} 

3D simulations show less pronounced boundaries of large-scale eddies, but the intermittent large current sheets are still present in the system \rev{with the weak guide field}. Figure \ref{fig:alignment3d}a and the linked \href{https://youtu.be/nY3F4bnTtEM}{video}\footnote{Direct link: \url{https://youtu.be/nY3F4bnTtEM} } demonstrate the distribution of the electric current $j_z$ in the planes perpendicular to the guide field, \rev{$B_z$}, at $t=1$, and Figure \ref{fig:alignment3d}b and the accompanying \href{https://youtu.be/8CRiWAZg_Bo}{video}\footnote{Direct link: \url{https://youtu.be/8CRiWAZg_Bo}} show the same at $t=2$. Similarly to the 2D results, intense current sheets occupy up to $4-5\%$ of the total volume of the domain\footnote{The larger filling fraction in 3D simulations is potentially attributed to the fact that for a similar value of resistivity, \mbox{$\eta=10^{-6}$}, the widths of current sheets are not fully converged.} and lead to $20\%$ of the total dissipation of the magnetic energy. 
Intermittent long current sheets are clearly plasmoid-unstable as shown by the insets in Figure \ref{fig:alignment3d}. A few initial eddies are still clearly seen at $t=1$, but many long intermittent current sheets are unaffected by the choice of the initial conditions.  At $t=2$ no visible features are associated with the initial conditions (see Figure \ref{fig:alignment3d}b).

\begin{figure}
\includegraphics[width=1.0\linewidth]{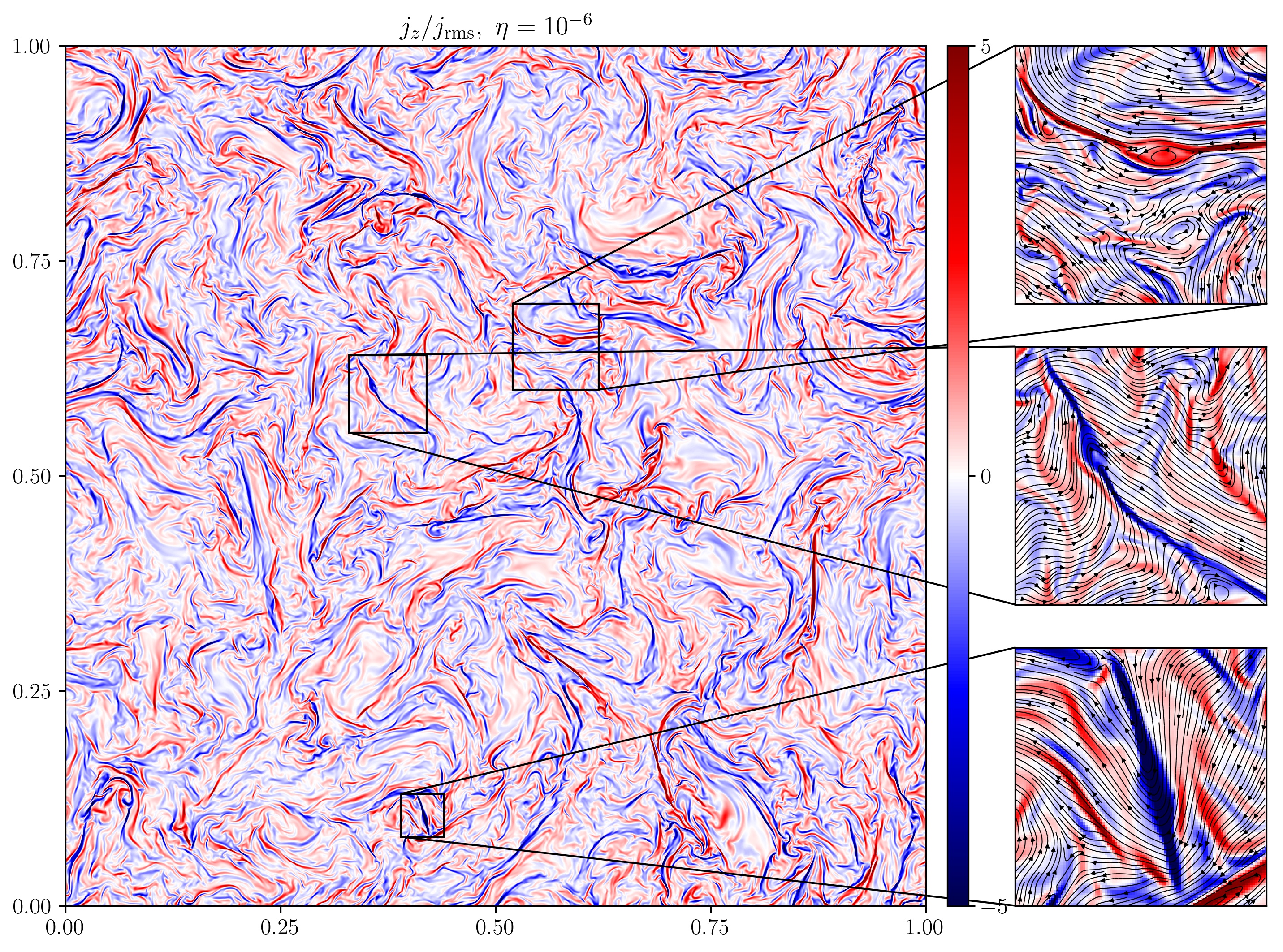}
\caption{\rev{3D SRRMHD simulation of highly-magnetized decaying turbulence, run 3D[d]. The color shows the out-of-plane component of the electric current $j_z$ in the snapshots at $t=2.0$, when $\langle\delta B\rangle/\langle B_z \rangle = 0.2$. The insets show zooms into individual current sheets which indicate plasmoid formation. The streamlines in the insets show \rev{the in-plane magnetic field lines. The current sheets in the middle and bottom insets do not show a perfect anti-parallel field geometry because the local guide field is tilted with respect to the plane of the snapshot}.}}
\label{fig:strongGF}
\end{figure}

\rev{Overall, the structure of the electric current in the 3D weak guide field simulation looks similar to the one in 2D (Figure \ref{fig:2d}a): there is a number of well-pronounced long, plasmoid unstable current sheets formed at the outer scale. Their formation is likely associated with the mergers and subsequent reconnection of large coherent structures \citep{Hosking}. Unlike in the weak guide field regime, the strong guide field simulation 3D[d] shows the statistical properties of ``aligned'' critically-balanced turbulence: the $k^{-3/2}_{\perp}$ spectrum and a pronounced dynamic alignment (dashed lines in Figure \ref{fig:alignment3d}c and d). The spatial distribution of the electric current is more uniform in this case (see Figure~\ref{fig:strongGF}). The absence of very long current sheets is consistent with the observation of a very few plasmoids in the simulation (see insets of Figure~\ref{fig:strongGF}). A possible explanation can \nrev{be} found in the small ratio of the length, $L_{\rm sheet}\sim 0.05$, for the sheets shown in the insets of Figure~\ref{fig:strongGF}, to the width of these sheets, which at our resolution is still limited by the numerical diffusion. We anticipate that the plasmoid instability can be more reliably captured at much higher spatial resolution: for the typical length, $L_{\rm sheet}\sim 0.05$, and the width-to-length ratio $\theta\approx 0.01$, one requires $(N_\delta/(L_{\rm sheet}\theta))^3 \approx 10000^3$ grid points, where $N_\delta\approx 5$ cells is the minimally desired resolution per width of the plasmoid-unstable current sheet.}

%\rev{The strong guide field simulation 3D[d] demonstrates less large-scale structured eddies and rather show many small scale eddies as depicted in Figure \ref{fig:strongGF}. This is consistent with flatter energy spectrum than in weak guide field simulation (Figure \ref{fig:alignment3d}c) meaning that smaller scale structures are more energetic and dynamically important. This more turbulent magnetic field obstructs formation of long intermittent current sheets emerging on the interface of colliding large eddies. Presented in the system small scale current sheets are supported against further compression by advected from upstream strong guide field. These short and thick current sheets can barely form plasmoids, a few of closed-to-formed plasmoids are shown in insets of Figure \ref{fig:strongGF}. As it is demonstrated in high resolution 2D simulations, active plasmoid formation still exists in high guide field system, but it  requires much higher resolution to resolve the onset of the instability. The needed to capture plasmoid chain formation numerical resolution is not achievable for modern computer resources.}

\begin{figure*}
\includegraphics[width=1.0\linewidth]{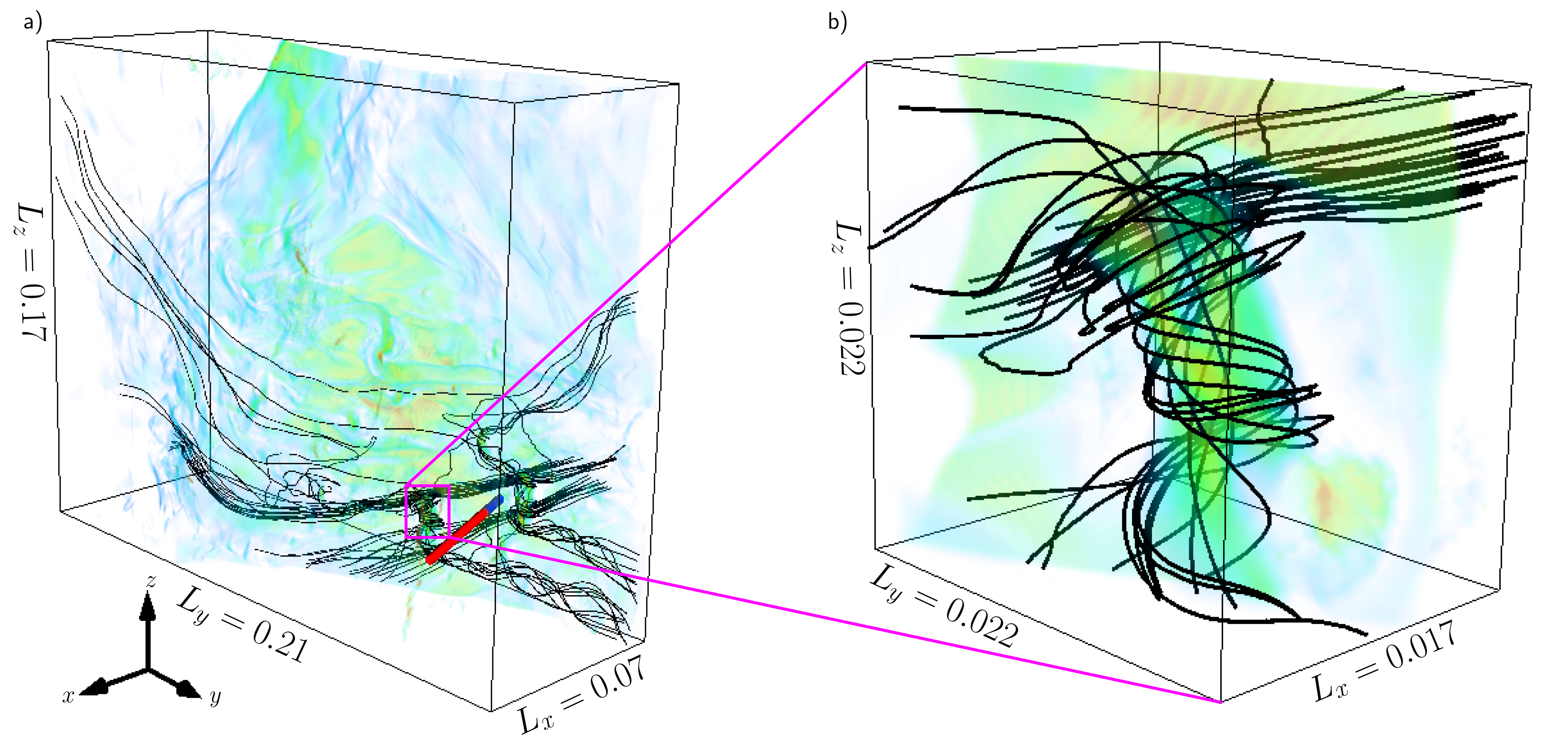}
\caption{3D volume rendering of the current density in a representative long current sheet in simulation 3D[a] at $t=1$. Color shows the amplitude of the current density, and thick black lines show magnetic field lines near plasmoids. a) Structure of a sheet. The red-blue line presents the slice across the current sheet shown in Figure \ref{fig:KH}, different colors of the line represent the different sides of the current sheet. b) Zoom into the structure of the plasmoid.}
\label{fig:3DCurrentSheet}
\end{figure*}

%\rev{this sentence is not needed already} In 3D, the current sheets have thick regions of enhanced current density, which we associate with plasmoids. 
The structure of a representative current sheet \rev{for the weak guide field simulation 3D[a]} is presented in Figure \ref{fig:3DCurrentSheet}. The volume render represents the current density amplitude, and solid black lines show selected magnetic field lines. The lower threshold for the volume rendering is chosen to be $\approx 2 j_{\rm rms}$, in order to remove the upstream regions without significant current. The initial (seed) points for the integration of magnetic field lines are set inside two randomly chosen plasmoids. Wrapped, helical magnetic field is responsible for the large current density inside the plasmoids. The helical structure allows longer plasmoids (or, flux tubes) to be kink-unstable if their length is large enough to exceed the Kruskal-Shafranov stability limit. This instability likely limits the life time of plasmoids in current sheets and their axial extension. A zoom into the 3D structure of a plasmoid is shown in Figure \ref{fig:3DCurrentSheet}b.

Acceleration of the flow from the X-point of a Sweet-Parker current sheet up to Alfv\'{e}nic speed creates a velocity shear, which may become unstable to the Kelvin-Helmholtz instability (KHI). The analytical non-relativistic stability criterion $\Delta u \lesssim v_A$ \citep{LoureiroKH} suggests that the strong upstream magnetic field can lead to the stabilization of the KHI for the velocity shear $\Delta u$ and the Alfv\'{e}n speed $v_A$ determined by the upstream magnetic field strength. A similar criterion was derived for a simplified model with ${\bf B} || {\bf v}$ in a fully relativistic case \citep{Osmanov}. \rev{Thus, we expect current sheets in highly-magnetized turbulence to be stabilized by the strong upstream magnetic field. To confirm this prediction, we conduct localized numerical experiments with conditions inferred from turbulence simulations (see Appendix \ref{Appendix:KH} for the description of the setups) that confirm that long plasmoid-unstable current sheets are Kelvin-Helmholtz-stable both in 2D and 3D simulations.}

\section{Conclusions} \label{sec:conclusion}

In this Letter we present the first 2D and 3D numerical SRRMHD simulations of highly magnetized  decaying turbulence. We calculate statistical properties of the turbulence, by analyzing a quasi steady state at two Alfv\'{e}n-crossing times of the simulation box. We show that the spectrum of magnetic energy in both cases is close to Boldyrev's spectrum, $k^{-3/2}_{\perp}$, and the \nrev{${\bf v}-{\bf B}$} dynamic alignment \rev{angle} follows an $l^{1/4}$ dependence. Despite the dynamic alignment \rev{angle of $\bf v$ and $\bf B$ fields} in 2D is perfectly following Boldyrev's prediction, \rev{its formation} cannot be explained by the uncertainty principle originally employed by \cite{Boldyrev_2006}. On the other hand, intermittent structures are vastly present in the simulations, favoring the theory of mutual shearing of Elsasser fields by \cite{Chandran_2015}\nrev{: an in-depth analysis of this approach is presented in Appendix \ref{Appendix:Int}}. %\rev{Also, matching of absence of $\delta {\bf z}_\pm$ dynamic alignment for weak guide field 3D simulation together with $-5/3$ the power spectrum slope indicates that $\delta {\bf z}_\pm$ alignment is responsible for reducing non-linearity of turbulence rather than ${\bf v}-{\bf b}$ one.} 
We demonstrate that long-lived intermittent current sheets form dynamically throughout the evolution. These sheets are plasmoid unstable and KH-stable. They occupy a very small fraction of the numerical domain but provide a significant fraction of the total magnetic field dissipation. 

In our simulations we only employ explicit resistivity while viscosity is dictated by the finite grid resolution. We expect that the magnetic energy dominates the kinetic energy at all scales, and dissipation is governed by resistivity. It will be useful to perform simulations with explicit viscosity and fixed magnetic Prandtl number ${\rm Pr_m}$ in the future studies, and to consider the trans-relativistic regime, $\sigma\sim1$. These studies can be applied to turbulence in the accretion disk-jet boundary with moderate magnetization \citep{Ripperda_2020}. 

In order to study the properties of intermittent current sheets in a statistical steady state, it is important to study driven turbulence in highly magnetized plasmas $\sigma\gg1$. High magnetization leads to efficient heating of the plasma due to the dissipation of magnetic energy and a significant drop of $\sigma$. To mediate the effect of runaway heating, radiative cooling of the plasma should be incorporated in the simulations \citep{Zhdankin_cooling}.

The limitation of computational resources does not allow to reach numerical resolutions significantly above $10000^3$ in the nearby future. This is too low to reach alignment angles substantially below $\theta\sim 0.1$ at the smallest scales. On the other hand, the intriguing similarity of statistical properties of 2D and 3D turbulence in our simulations makes it interesting to perform even higher resolution simulations of 2D turbulence. The most significant milestone will be a resolution of $\sim (10^8)^2$ which allows progressively elongated eddies to reach an alignment angle $\theta\sim0.01$ corresponding to the plasmoid instability of these eddies. The steepening of the turbulence spectrum due to the onset of the plasmoid instability in intermittent current sheets (or due to the linear tearing instability in elongated eddies, \citealt{Boldyrev_Loureiro}) can be measured reliably at resolutions of $\sim (10^6)^2$, realistically attainable in the nearby future, in particular with the AMR criterion we propose here.

\section{Acknowledgements}

We acknowledge useful discussions with Lev Arzamasskiy, Amitava Bhattacharjee, Benjamin Chandran, Luca Comisso, Mikhail Medvedev, Joonas N\"{a}ttil\"{a}, Jason TenBarge, James Stone, and help in navigating through 3D-visualization by Hayk Hakobyan. \rev{The authors acknowledge insightful comments by the anonymous referee which helped to significantly improve the manuscript.} The computational resources and services used in this work were provided by facilities supported by the Scientific Computing Core at the Flatiron Institute, a division of the Simons Foundation; and by the VSC (Flemish Supercomputer Center), funded by the Research Foundation Flanders (FWO) and the Flemish Government – department EWI. 
A.C. gratefully acknowledges support and hospitality from the Simons Foundation through the pre-doctoral program at the Center for
Computational Astrophysics, Flatiron Institute. B.R. is supported by a Joint Princeton/Fellowship Postdoctoral Fellowship. A.P. acknowledges support by the National Science Foundation under Grant No. AST-1910248.

\rev{\textit{Software}: {\tt{BHAC}}, \citep{BHAC1, BHAC2,resBHAC}, Python \citep{python1, python2}, NumPy \citep{NumPy}, Matplotlib \citep{matplotlib}, Mayavi \citep{mayavi}}

\appendix

\section{Adaptive Mesh Refinement and Convergence Study}\label{Appendix:AMR}

\begin{figure*}
\includegraphics[width=1.0\linewidth]{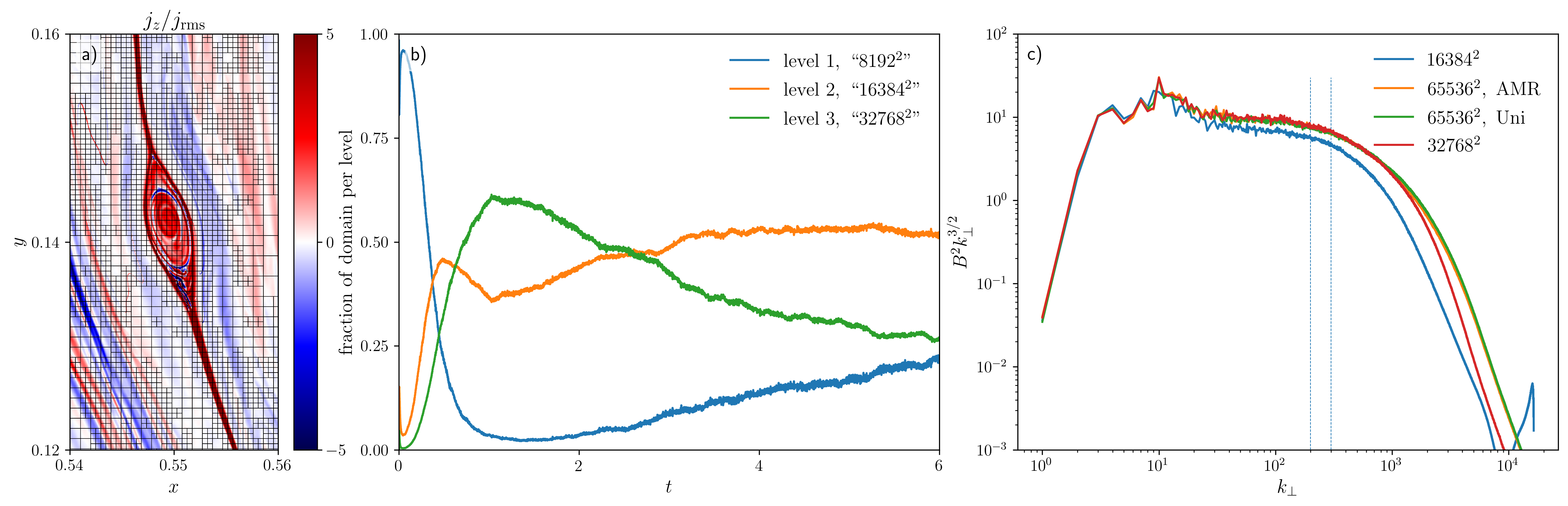}
\caption{Analysis of numerical convergence of 2D simualations with numerical resolution. a) An example of the structure of the refined grid close to a plasmoid-unstable current sheet. b) Coverage of the numerical domain by blocks of different refinement levels during the simulations. c) Resolution study for 2D simulations with uniform  $16384^2, 32768^2, 65536^2 $ and AMR $65536^2$, which shows the comparison of magnetic energy spectra. Vertical dashed lines show the end of the inertial range for simulations with $16384^2$ (left line) and $32768^2/65536^2$ (right line) grid points. }
\label{fig:AMR}
\end{figure*}

For the fully resolved and converged 2D simulations we present the adaptive mesh refinement (AMR) criterion we designed to accelerate the simulations and to simultaneously capture the main properties of the turbulence and dissipative structures. The main principle is that the largest eddies are resolved by many cells at low resolution. To capture the physics at smallest scales, one needs to refine the resolution in the smallest eddies, capturing both velocity and magnetic field gradients. We define the characteristic sizes of the eddies as 
\begin{equation}
    l_{\bf v} = \frac{|\left(\nabla \times {\bf v}\right)_z|}{\sqrt{v_x^2+v_y^2}},~~~l_{\bf B} = \frac{|\left(\nabla \times {\bf B}\right)_z|}{\sqrt{B_x^2+B_y^2}}.
\end{equation}

The refinement routine is called if the size of \textit{any} of the two quantities is less than a threshold value: $l_{{\bf v}, {\bf B}} < \alpha \Delta x$ at the point. Coefficient $\alpha$ is chosen to be such that the threshold scale is larger than the numerical resistive scale. In the simulations we use $\alpha=8$, which is larger than the numerical resistive scale in simulations 2D[a] and 2D[c], and $\Delta x$ is the grid spacing at a given resolution. Coarsening of the grid in the numerical domain is only allowed if \textit{both} quantities at a given  grid point are larger than the threshold. Since the electric current density is roughly given by the gradient of the magnetic field, ${\bf j}\sim\nabla \times {\bf B}$, regions of the large electric current density (indicating current sheets) are automatically refined. Since the inverse cascade is very pronounced in 2D simulations, AMR shows very high efficiency at early times, when the spectrum is being formed, and at later times, when small eddies merge in larger ones (see Figure \ref{fig:AMR}a for $\eta=10^{-6}$). Since the resistive scale is much larger for $\eta=10^{-5}$, the coverage by the highest resolution level does not exceed $15\%$ in this case. 

The threshold value is tested for a resolution of $32768^2$ grid points by comparing spectra of the magnetic field energy for uniform grid and AMR-enabled runs (where the effective resolution for the AMR runs indicates the total resolution if the whole domain were refined to the highest AMR level) at the same moment in time (see Figure \ref{fig:AMR}b). Interestingly, the most accurate spectra are produced by simulations where the refinement algorithm is called only every $50-100$ time-steps, most likely due to less numerical noise being generated during the refining and coarsening of the grid and re-interpolation. The frequency of the refinement calls is defined to ensure that the finest structures are always located inside the refined grid block during their motion in the domain. For the bulk velocity of the fluid $u\approx0.1c$, and CFL number $0.4$, an element of the fluid travels about $20$ cells between two calls of the refinement, while the minimum size of a refined grid is $32^2$ cells. 

This AMR strategy does not seem to be effective in 3D simulations due to the overall low grid resolution, compared to the extreme resolutions employed in 2D: AMR automatically chooses the resolution needed to resolve all the features in the block. Since the size of all features in the flow is rather defined by the numerical resolution than by an explicit resistivity, AMR refines the whole domain up to the highest available resolution. It is impossible to find a reasonable threshold $\alpha$ for the 2D counterpart of the highest resolution 3D run with $3200^3$ grid points: any chosen $\alpha$ either truncates the inertial range of the spectrum or refines the whole domain shortly after the start of the simulation. Figure \ref{fig:AMR}c  demonstrates that a base resolution of $\gtrsim 32000^2$ grid points is needed to fully resolve the resistive scale for $\eta=10^{-6}$ and keep the inertial range of the turbulence unaffected by the resolution.  In order to demonstrate this, we compare spectra for resolutions with $16384^2,~32768^2,~65536^2$ points. 

\section{Kelvin-Helmholtz Stability of Current Sheets}\label{Appendix:KH}

\begin{figure*}
\includegraphics[width=1.0\linewidth]{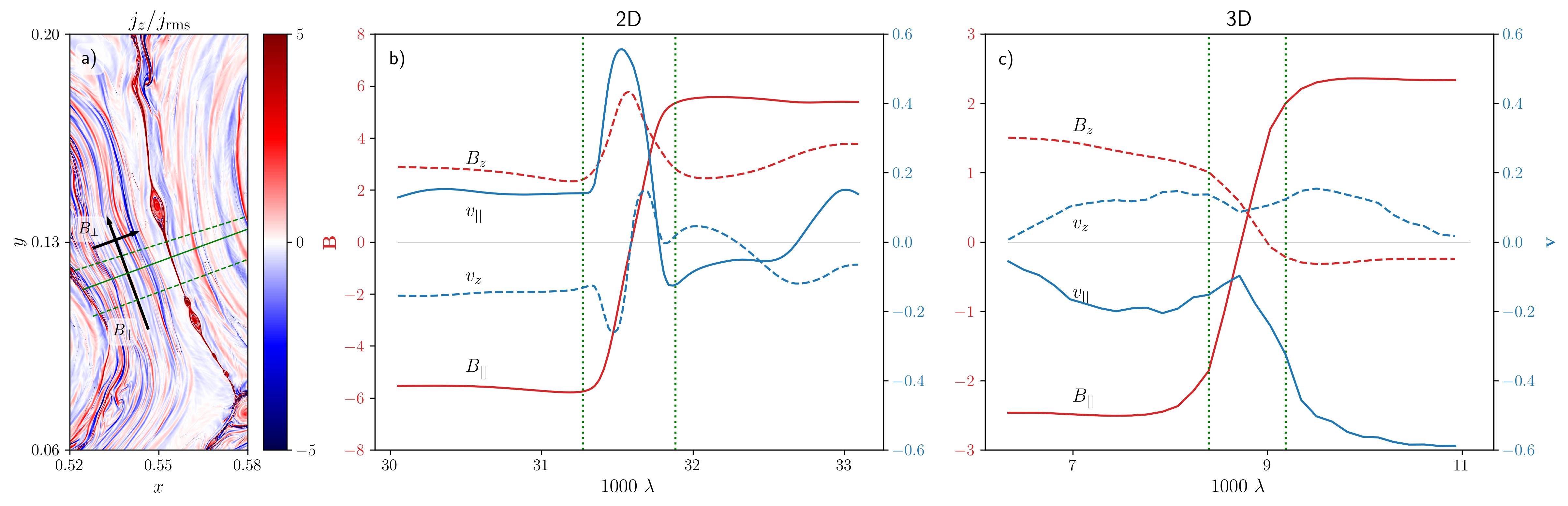}
\caption{Analysis of the KH-stability of current sheets in turbulence. a) Slices across the current sheet in simulation 2D[a], used to extract the shear flow parameters. \rev{Arrows show directions of $B_{||},B_{\perp}$, or $v_{||}, v_\perp$}. b) Behavior of the reconnecting in-plane and out-of-plane magnetic field components and parallel velocity along the slice shown with a solid line in panel a. c) Similar quantities along the slice across the current sheet in 3D simulation 3D[a].}
\label{fig:KH}
\end{figure*}

In order to study the stability of the magnetized shear flow in  our plasmoid-unstable current sheets \rev{in 2D}, we calculate the value of the in-plane reconnecting magnetic field component $B_{||}$ and the out-of-plane component $B_z$ as well as velocity field components $v_{||}, v_z$ for each of the three slices shown in Figure~\ref{fig:KH}a \rev{by green lines}. For all these slices we find $|{\bf B}_{\perp}| \ll |{\bf B}_{||}|$ and $|{\bf v}_{\perp}| \ll |{\bf v}_{||}|$, where $||$ represents the direction parallel to the current sheet at a given point in the slice \rev{(the arrows in Figure~\ref{fig:KH}a indicate parallel and perpendicular directions, and the $z$-direction is out-of-plane)}. We show the typical behavior of these parameters in Figure~\ref{fig:KH}b, which implies that the flow satisfies the non-relativistic stability criterion \rev{$|\delta v| < |\delta B|$}. 

For each slice across the current sheet, we run a local simulation of the shear flow, with one flow having parameters ($\rho, B_{||}, B_z, v_{||}, v_z$) given by the upstream of the current sheet, and its counter-flow having parameters from the interior of the current sheet, \rev{particularly,} $B_{||}=0$. Zero parallel magnetic field in one of the two interacting flows prohibits reconnection at their interface in these experiments, but allows to study the KHI. Plasma pressure is adjusted to maintain the force balance across the interface of the flows. %\rev{We are interested in the current  one with larger velocity shear $\delta v_{||}$.}

We run simulations with a resolution of $4096^2$ grid points for 20 light-crossing times along the sheet, which exceeds the life time of intermittent current sheets in the full 2D turbulence simulation. For the intermediate slice (shown by the solid line in Figure \ref{fig:KH}a), we run an AMR-enabled simulation with an effective resolution of $32768^2$ grid points, which resolves all the scales up to the resistive scale for a resistivity $\eta=10^{-6}$(see Appendix \ref{Appendix:AMR} for the resolution study). \rev{This finest grid covers the whole interface of the flows at any moment of the simulation.} In all of these experiments we do not observe any instability growth. This implies that the in-plane magnetic field in the upstream of the current sheet is capable of preserving KH-stability in both the upstream and downstream of the current sheet.

\begin{figure*}%[hbt!]
\includegraphics[width=1.0\linewidth]{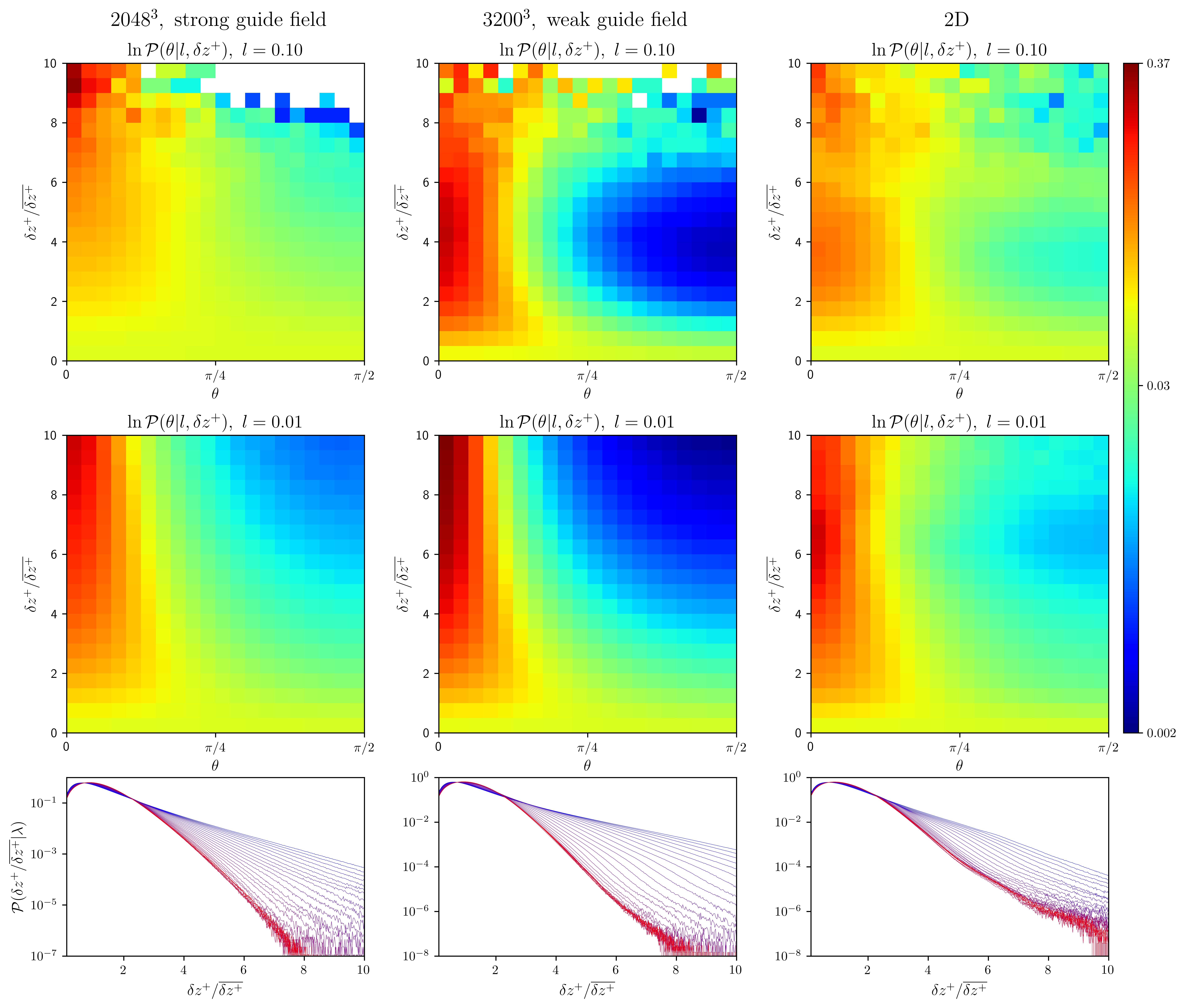}
\caption{\nrev{Intermittency of the turbulence: The two top rows show the conditional probability distribution function (PDF) of the dynamic alignment angle for a fixed amplitude of the Elsasser field $\delta z^+/\overline{\delta z^+_l}$  and at a fixed scale $l$ (the top row, $l=0.1$, corresponds to the low values of $k_\perp$, approximately at the beginning of the intertial range; the middle row, $l=0.01$,- corresponds to the high-$k_\perp$  end of the inertial range), measured at $t=2$. Here, $\overline{\delta z^+_l}=\exp{\langle \ln{\delta z^+ | l}\rangle}$ is the geometrical mean of $\delta z^+$ at a fixed scale $l$. The bottom row shows the PDF of $\delta z^+/\overline{\delta z^+_l}$ for 30 logarithmically distributed scales between $l=0.9L$ (red line) to $l=0.01L$ (blue line).}}
\label{fig:intermittency}
\end{figure*}

In order to explore the KH-stability of the 3D current sheet, we select slices between the two plasmoids (red-blue line in Figure \ref{fig:3DCurrentSheet}), and perform test simulations in a 3D setup with a geometry similar to the one described above for 2D simulations. \rev{This slice is also normal to the surface of the current sheet at the point of their intersection.} We test the sheet's stability by running a 3D simulation with a resolution of $1024^3$ grid points for 20 light-crossing times along the shear interface, with parameters corresponding to the slice shown in Figure \ref{fig:3DCurrentSheet}c. 

\section{Amplitude-dependent intermittency of the turbulence.}\label{Appendix:Int}
\nrev{Since the lack of scale invariance is the most prominent sign of intermittency, we focus on the scale dependence of the Elsasser field increments, $\delta z^+$. The results presented in this Appendix are similar for $\delta z^-$ as we expect the turbulence to be balanced, $\delta z^+ \sim \delta z^-$. As shown by \cite{Mallet2015}, the scale invariance can be characterized by the similarity of the conditional probability distribution functions (PDFs) $\mathcal{P}(\delta z^+ | l)$ of $\delta z^+$ at different scales $l$. We computed these PDFs for 2D (run 2D[a]) and 3D (runs 3D[a] and 3D[d]) simulations. To measure the PDF, we use a set of 30 logarithmically spaced scales $\{l_i\}$ from $0.01L$ to $0.9L$. The smallest scale $l_{1}$, corresponding to $k_\perp\approx100$, lies deeply in the inertial range of the energy spectrum, while the largest scale of the considered set, $l_{30}$, is located at the outer scale of the turbulence. As the bottom row of Figure \ref{fig:intermittency} shows, the smaller-scale eddies (darker lines) have higher probability to reach large normalized amplitudes of $\delta z^+$. The flattening of the PDFs at smaller scales can be attributed to the sheet-like structures emerging at these scales. For 2D and weak guide field 3D simulations, the presence of long current sheets can also explain a flatter tail of the PDF at the larger scales, while the PDF for the strong guide field 3D case has an abrupt cutoff at the high $\delta z^+$ for the same eddy sizes. We normalized $\delta z^+(l)$ by a geometrical mean of $\overline{\delta z^+}=\exp{\langle \ln{\delta z^+ |l} \rangle}$ at a given scale $l$, as it is less sensitive to outliers than an arithmetical mean.}

\nrev{The intermittent, scale-dependent, nature of the dynamic alignment can also be shown by measuring the PDF of the dynamic alignment angle at given scales, as considered by \cite{Dong2018}. We are, however, also interested in testing the assumption of \cite{Chandran_2015} that large $\delta z^+$ rotates $\delta z^-$ into alignment, while balanced collisions $\delta z^+ \sim \delta z^- \sim \overline{\delta z^{\pm}}$ are not aligned. This anti-correlation of the alignment angle with the amplitude of $\delta z^+$ contradicts the intuitive explanation of the dynamic alignment by an uncertainty principle. To test this, we measure the conditional PDF of the dynamic alignment angle $\mathcal{P}(\theta | l, \delta z^+)$ for a given scale $l$ and the amplitude of the Elsasser field $\delta z^+/ \overline{\delta z^+}$. The middle and top rows of Figure \ref{fig:intermittency} show that the the prediction is matched perfectly for strong guide field 3D turbulence: the larger $\delta z^+$, the more aligned $\delta z^+$ and  $\delta z^-$ are. For 2D and weak guide field 3D turbulence there is a deviation from this prediction at the outer scale: while the statement holds for intermediate amplitudes of $\delta z^+$, at high amplitudes eddies become uncorrelated again. The most powerful increments $\delta z^+$ are associated to current sheets and plasmoids, and one can expect that circular plasmoids have an alignment angle (the ratio of two length scales of the eddy) $\theta\sim1$ that can explain decorrelation of the alignment angle at high $\delta z^+/\overline{\delta z^+}$.}
%It is worth noting that the deviation from the log-normal-like PDF of $\delta z^+$ at the outer scale happens at the same $\delta z^+$ as the PDF of the dynamic alignment angle becomes more uniform (or, eddies become interacting in more coupled and non-linear way).} 

\bibliography{sample63}{}

\begin{thebibliography}{}
\expandafter\ifx\csname natexlab\endcsname\relax\def\natexlab#1{#1}\fi
\providecommand{\url}[1]{\href{#1}{#1}}
\providecommand{\dodoi}[1]{doi:~\href{http://doi.org/#1}{\nolinkurl{#1}}}
\providecommand{\doeprint}[1]{\href{http://ascl.net/#1}{\nolinkurl{http://ascl.net/#1}}}
\providecommand{\doarXiv}[1]{\href{https://arxiv.org/abs/#1}{\nolinkurl{https://arxiv.org/abs/#1}}}

\bibitem[{{Begelman}(1998)}]{Begelman1998}
{Begelman}, M.~C. 1998, \apj, 493, 291.
\newblock \doarXiv{astro-ph/9708142}

\bibitem[{{Beloborodov}(2017)}]{Beloborodov17}
{Beloborodov}, A.~M. 2017, \apj, 850, 141, \dodoi{10.3847/1538-4357/aa8f4f}

\bibitem[{Beloborodov(2020)}]{beloborodov2020emission}
Beloborodov, A.~M. 2020, Emission of magnetar bursts and precursors of neutron
  star mergers.
\newblock \doarXiv{2011.07310}

\bibitem[{{Bhattacharjee} {et~al.}(2009){Bhattacharjee}, {Huang}, {Yang}, \&
  {Rogers}}]{bhattacharjee2009}
{Bhattacharjee}, A., {Huang}, Y.-M., {Yang}, H., \& {Rogers}, B. 2009, Physics
  of Plasmas, 16, 112102

\bibitem[{{Boldyrev}(2005)}]{Boldyrev_2005}
{Boldyrev}, S. 2005, \apjl, 626, L37.
\newblock \doarXiv{astro-ph/0503053}

\bibitem[{{Boldyrev}(2006)}]{Boldyrev_2006}
---. 2006, \prl, 96, 115002.
\newblock \doarXiv{astro-ph/0511290}

\bibitem[{{Boldyrev} \& {Loureiro}(2017)}]{Boldyrev_Loureiro}
{Boldyrev}, S., \& {Loureiro}, N.~F. 2017, \apj, 844, 125.
\newblock \doarXiv{1706.07139}

\bibitem[{{Cerutti} {et~al.}(2015){Cerutti}, {Philippov}, {Parfrey}, \&
  {Spitkovsky}}]{Cerutti2015}
{Cerutti}, B., {Philippov}, A., {Parfrey}, K., \& {Spitkovsky}, A. 2015,
  \mnras, 448, 606, \dodoi{10.1093/mnras/stv042}

\bibitem[{{Chandran} {et~al.}(2018){Chandran}, {Foucart}, \&
  {Tchekhovskoy}}]{Chandran_2018}
{Chandran}, B. D.~G., {Foucart}, F., \& {Tchekhovskoy}, A. 2018, Journal of
  Plasma Physics, 84, 905840310.
\newblock \doarXiv{1707.06216}

\bibitem[{{Chandran} {et~al.}(2015){Chandran}, {Schekochihin}, \&
  {Mallet}}]{Chandran_2015}
{Chandran}, B.~D.~G., {Schekochihin}, A.~A., \& {Mallet}, A. 2015, \apj, 807,
  39.
\newblock \doarXiv{1403.6354}

\bibitem[{{Cho} \& {Vishniac}(2000)}]{Cho_Vishniac}
{Cho}, J., \& {Vishniac}, E.~T. 2000, \apj, 539, 273, \dodoi{10.1086/309213}

\bibitem[{{Comisso} \& {Sironi}(2018)}]{Comisso1}
{Comisso}, L., \& {Sironi}, L. 2018, \prl, 121, 255101.
\newblock \doarXiv{1809.01168}

\bibitem[{{Del Zanna} {et~al.}(2016){Del Zanna}, {Papini}, {Landi}, {Bugli}, \&
  {Bucciantini}}]{DelZanna}
{Del Zanna}, L., {Papini}, E., {Landi}, S., {Bugli}, M., \& {Bucciantini}, N.
  2016, \mnras, 460, 3753, \dodoi{10.1093/mnras/stw1242}

\bibitem[{{Dong} {et~al.}(2018){Dong}, {Wang}, {Huang}, {Comisso}, \&
  {Bhattacharjee}}]{Dong2018}
{Dong}, C., {Wang}, L., {Huang}, Y.-M., {Comisso}, L., \& {Bhattacharjee}, A.
  2018, \prl, 121, 165101.
\newblock \doarXiv{1804.07361}

\bibitem[{{Elsasser}(1950)}]{Elsasser:1950}
{Elsasser}, W.~M. 1950, Physical Review, 79, 183,
  \dodoi{10.1103/PhysRev.79.183}

\bibitem[{Frisch(1995)}]{frisch_1995}
Frisch, U. 1995, Turbulence: The Legacy of A. N. Kolmogorov (Cambridge
  University Press), \dodoi{10.1017/CBO9781139170666}

\bibitem[{{Goldreich} \& {Sridhar}(1995)}]{GS95}
{Goldreich}, P., \& {Sridhar}, S. 1995, \apj, 438, 763

\bibitem[{{Goldreich} \& {Sridhar}(1997)}]{GS_revisited}
---. 1997, \apj, 485, 680.
\newblock \doarXiv{astro-ph/9612243}

\bibitem[{Guo {et~al.}(2014)Guo, Li, Daughton, \& Liu}]{Guo_2014}
Guo, F., Li, H., Daughton, W., \& Liu, Y.-H. 2014, \prl, 113

\bibitem[{{Hosking} \& {Schekochihin}(2020)}]{Hosking}
{Hosking}, D.~N., \& {Schekochihin}, A.~A. 2020, arXiv e-prints,
  arXiv:2012.01393.
\newblock \doarXiv{2012.01393}

\bibitem[{Hunter(2007)}]{matplotlib}
Hunter, J.~D. 2007, Computing in Science Engineering, 9, 90,
  \dodoi{10.1109/MCSE.2007.55}

\bibitem[{{Iroshnikov}(1963)}]{Iroshnikov1963}
{Iroshnikov}, P.~S. 1963, \azh, 40, 742

\bibitem[{{Kraichnan}(1965)}]{Kraichnan1965}
{Kraichnan}, R.~H. 1965, Physics of Fluids, 8, 1385

\bibitem[{{Loureiro} {et~al.}(2007){Loureiro}, {Schekochihin}, \&
  {Cowley}}]{Loureiro2007}
{Loureiro}, N.~F., {Schekochihin}, A.~A., \& {Cowley}, S.~C. 2007, Physics of
  Plasmas, 14, 100703.
\newblock \doarXiv{astro-ph/0703631}

\bibitem[{{Loureiro} {et~al.}(2013){Loureiro}, {Schekochihin}, \&
  {Uzdensky}}]{LoureiroKH}
{Loureiro}, N.~F., {Schekochihin}, A.~A., \& {Uzdensky}, D.~A. 2013, \pre, 87,
  013102.
\newblock \doarXiv{1208.0966}

\bibitem[{{Loureiro} {et~al.}(2009){Loureiro}, {Uzdensky}, {Schekochihin},
  {Cowley}, \& {Yousef}}]{Loureiro_2009}
{Loureiro}, N.~F., {Uzdensky}, D.~A., {Schekochihin}, A.~A., {Cowley}, S.~C.,
  \& {Yousef}, T.~A. 2009, \mnras, 399, L146,
  \dodoi{10.1111/j.1745-3933.2009.00742.x}

\bibitem[{{Lyubarsky}(1992)}]{Lyubarsky1992}
{Lyubarsky}, Y.~E. 1992, Soviet Astronomy Letters, 18, 356

\bibitem[{{Lyubarsky}(2005)}]{Lyubarsky2005}
---. 2005, \mnras, 358, 113, \dodoi{10.1111/j.1365-2966.2005.08767.x}

\bibitem[{{Mahlmann} {et~al.}(2020){Mahlmann}, {Levinson}, \&
  {Aloy}}]{Mahlmann_2020}
{Mahlmann}, J.~F., {Levinson}, A., \& {Aloy}, M.~A. 2020, \mnras, 494, 4203.
\newblock \doarXiv{2001.03171}

\bibitem[{{Mallet} {et~al.}(2015){Mallet}, {Schekochihin}, \&
  {Chandran}}]{Mallet2015}
{Mallet}, A., {Schekochihin}, A.~A., \& {Chandran}, B.~D.~G. 2015, \mnras, 449,
  L77, \dodoi{10.1093/mnrasl/slv021}

\bibitem[{{Mallet} {et~al.}(2017){Mallet}, {Schekochihin}, \&
  {Chandran}}]{Mallet_2017}
---. 2017, \mnras, 468, 4862.
\newblock \doarXiv{1612.07604}

\bibitem[{{Mallet} {et~al.}(2016){Mallet}, {Schekochihin}, {Chandran}, {Chen},
  {Horbury}, {Wicks}, \& {Greenan}}]{Mallet2016}
{Mallet}, A., {Schekochihin}, A.~A., {Chandran}, B.~D.~G., {et~al.} 2016,
  \mnras, 459, 2130, \dodoi{10.1093/mnras/stw802}

\bibitem[{{Mason} {et~al.}(2006){Mason}, {Cattaneo}, \&
  {Boldyrev}}]{Mason_2006}
{Mason}, J., {Cattaneo}, F., \& {Boldyrev}, S. 2006, \prl, 97, 255002.
\newblock \doarXiv{astro-ph/0602382}

\bibitem[{{Millman} \& {Aivazis}(2011)}]{python2}
{Millman}, K., \& {Aivazis}, M. 2011, Computing in Science \& Engineering, 13,
  9 , \dodoi{10.1109/MCSE.2011.36}

\bibitem[{{N{\"a}ttil{\"a}} \& {Beloborodov}(2020)}]{Nattila}
{N{\"a}ttil{\"a}}, J., \& {Beloborodov}, A.~M. 2020, arXiv e-prints,
  arXiv:2012.03043.
\newblock \doarXiv{2012.03043}

\bibitem[{Oliphant(2007)}]{python1}
Oliphant, T.~E. 2007, Computing in Science Engineering, 9, 10,
  \dodoi{10.1109/MCSE.2007.58}

\bibitem[{{Olivares} {et~al.}(2019){Olivares}, {Porth}, {Davelaar}, {Most},
  {Fromm}, {Mizuno}, {Younsi}, \& {Rezzolla}}]{BHAC2}
{Olivares}, H., {Porth}, O., {Davelaar}, J., {et~al.} 2019, \aap, 629, A61.
\newblock \doarXiv{1906.10795}

\bibitem[{{Osmanov} {et~al.}(2008){Osmanov}, {Mignone}, {Massaglia}, {Bodo}, \&
  {Ferrari}}]{Osmanov}
{Osmanov}, Z., {Mignone}, A., {Massaglia}, S., {Bodo}, G., \& {Ferrari}, A.
  2008, \aap, 490, 493.
\newblock \doarXiv{0802.2607}

\bibitem[{{Parker}(1957)}]{parker_1957}
{Parker}, E.~N. 1957, \jgr, 62, 509, \dodoi{10.1029/JZ062i004p00509}

\bibitem[{{Perez} {et~al.}(2012){Perez}, {Mason}, {Boldyrev}, \&
  {Cattaneo}}]{Perez_2012}
{Perez}, J.~C., {Mason}, J., {Boldyrev}, S., \& {Cattaneo}, F. 2012, Physical
  Review X, 2, 041005.
\newblock \doarXiv{1209.2011}

\bibitem[{{Porth} {et~al.}(2017){Porth}, {Olivares}, {Mizuno}, {Younsi},
  {Rezzolla}, {Moscibrodzka}, {Falcke}, \& {Kramer}}]{BHAC1}
{Porth}, O., {Olivares}, H., {Mizuno}, Y., {et~al.} 2017, Computational
  Astrophysics and Cosmology, 4, 1.
\newblock \doarXiv{1611.09720}

\bibitem[{{Ramachandran} \& {Varoquaux}(2011)}]{mayavi}
{Ramachandran}, P., \& {Varoquaux}, G. 2011, Computing in Science \&
  Engineering, 13, 40 , \dodoi{10.1109/MCSE.2011.35}

\bibitem[{{Ripperda} {et~al.}(2020){Ripperda}, {Bacchini}, \&
  {Philippov}}]{Ripperda_2020}
{Ripperda}, B., {Bacchini}, F., \& {Philippov}, A.~A. 2020, \apj, 900, 100.
\newblock \doarXiv{2003.04330}

\bibitem[{{Ripperda} {et~al.}(2021){Ripperda}, {Liska}, {Chatterjee}, {Musoke},
  {Philippov}, {Markoff}, {Tchekhovskoy}, \& {Younsi}}]{RipperdaFlares}
{Ripperda}, B., {Liska}, M., {Chatterjee}, K., {et~al.} 2021, arXiv e-prints,
  arXiv:2109.15115.
\newblock \doarXiv{2109.15115}

\bibitem[{Ripperda {et~al.}(2019)Ripperda, Porth, Sironi, \&
  Keppens}]{Ripperda:2019a}
Ripperda, B., Porth, O., Sironi, L., \& Keppens, R. 2019, Monthly Notices of
  the Royal Astronomical Society, 485, 299–314.
\newblock \url{http://dx.doi.org/10.1093/mnras/stz387}

\bibitem[{{Ripperda} {et~al.}(2019){Ripperda}, {Bacchini}, {Porth}, {Most},
  {Olivares}, {Nathanail}, {Rezzolla}, {Teunissen}, \& {Keppens}}]{resBHAC}
{Ripperda}, B., {Bacchini}, F., {Porth}, O., {et~al.} 2019, \apjs, 244, 10.
\newblock \doarXiv{1907.07197}

\bibitem[{{Sironi} \& {Spitkovsky}(2014)}]{Sironi_Spitkovsky}
{Sironi}, L., \& {Spitkovsky}, A. 2014, \apjl, 783, L21.
\newblock \doarXiv{1401.5471}

\bibitem[{{Sweet}(1958)}]{sweet_1958}
{Sweet}, P.~A. 1958, in Electromagnetic Phenomena in Cosmical Physics, ed.
  B.~{Lehnert}, Vol.~6, 123

\bibitem[{{Takamoto} \& {Lazarian}(2017)}]{Takamoto_2017}
{Takamoto}, M., \& {Lazarian}, A. 2017, \mnras, 472, 4542,
  \dodoi{10.1093/mnras/stx2292}

\bibitem[{{TenBarge} {et~al.}(2021){TenBarge}, {Ripperda}, {Chernoglazov},
  {Bhattacharjee}, {Mahlmann}, {Most}, {Juno}, {Yuan}, \&
  {Philippov}}]{TenBarge2021}
{TenBarge}, J.~M., {Ripperda}, B., {Chernoglazov}, A., {et~al.} 2021, arXiv
  e-prints, arXiv:2105.01146.
\newblock \doarXiv{2105.01146}

\bibitem[{{Uzdensky} {et~al.}(2010){Uzdensky}, {Loureiro}, \&
  {Schekochihin}}]{uzdensky2010}
{Uzdensky}, D.~A., {Loureiro}, N.~F., \& {Schekochihin}, A.~A. 2010, \prl, 105,
  235002, \dodoi{10.1103/PhysRevLett.105.235002}

\bibitem[{{van der Walt} {et~al.}(2011){van der Walt}, {Colbert}, \&
  {Varoquaux}}]{NumPy}
{van der Walt}, S., {Colbert}, S.~C., \& {Varoquaux}, G. 2011, Computing in
  Science and Engineering, 13, 22, \dodoi{10.1109/MCSE.2011.37}

\bibitem[{Werner {et~al.}(2015)Werner, Uzdensky, Cerutti, Nalewajko, \&
  Begelman}]{Werner_2015}
Werner, G.~R., Uzdensky, D.~A., Cerutti, B., Nalewajko, K., \& Begelman, M.~C.
  2015, \apj, 816, L8

\bibitem[{{Zhdankin} {et~al.}(2021){Zhdankin}, {Uzdensky}, \&
  {Kunz}}]{Zhdankin_cooling}
{Zhdankin}, V., {Uzdensky}, D.~A., \& {Kunz}, M.~W. 2021, \apj, 908, 71.
\newblock \doarXiv{2007.12050}

\bibitem[{{Zhdankin} {et~al.}(2013){Zhdankin}, {Uzdensky}, {Perez}, \&
  {Boldyrev}}]{Zhdankin}
{Zhdankin}, V., {Uzdensky}, D.~A., {Perez}, J.~C., \& {Boldyrev}, S. 2013,
  \apj, 771, 124.
\newblock \doarXiv{1302.1460}

\bibitem[{{Zhdankin} {et~al.}(2017){Zhdankin}, {Werner}, {Uzdensky}, \&
  {Begelman}}]{Zhdankin17}
{Zhdankin}, V., {Werner}, G.~R., {Uzdensky}, D.~A., \& {Begelman}, M.~C. 2017,
  \prl, 118, 055103, \dodoi{10.1103/PhysRevLett.118.055103}

\bibitem[{{Zhou} {et~al.}(2020){Zhou}, {Loureiro}, \& {Uzdensky}}]{Zhou_2020}
{Zhou}, M., {Loureiro}, N.~F., \& {Uzdensky}, D.~A. 2020, Journal of Plasma
  Physics, 86, 535860401.
\newblock \doarXiv{2001.07291}

\bibitem[{{Zrake} \& {MacFadyen}(2012)}]{Zrake_2012}
{Zrake}, J., \& {MacFadyen}, A.~I. 2012, \apj, 744, 32.
\newblock \doarXiv{1108.1991}

\end{thebibliography}
\bibliographystyle{aasjournal}

\end{document}